\documentclass[12pt,preprint,numberedappendix]{aastex}
\usepackage{graphicx}
\usepackage{natbib}
\usepackage{aas_macros}
\usepackage[section]{placeins}
\usepackage{subfigure}
\usepackage{float}
\usepackage{color}

\usepackage[scaled]{helvet}

\usepackage[T1]{fontenc}

\def\h2{${\rm\,H_2}$}

\makeatletter 

\def\pc{{\rm\,pc}}

\def\vol#1  {{{#1}{\rm,}\ }}

\newcount\refno
\newcount\rfno
\def\eq{$^{\the\refno\ }$\advance\refno by 1}
\def\ad{\advance\rfno by 1}

\def\clock{\count0=\time \divide\count0 by 60
     \count1=\count0 \multiply\count1 by -60 \advance\count1
     by \time 
     \number\count0:\ifnum\count1<10{0\number\count1}
     \else\number\count1\fi}

\def\myputfigure#1#2#3#4#5%
{\hskip0.03\textwidth\vskip#5pt%\hfill
\makebox[0pt]{\hskip#2in
\includegraphics[width=#3\textwidth]{#1}}\vskip#4pt\hfill}

\newcount\refno
\newcount\rfno
\def\eq{$^{\the\refno\ }$\advance\refno by 1}
\def\ad{\advance\rfno by 1}

\definecolor{burntorange}{rgb}{1,0.4,0.2}

\definecolor{burntorange}{rgb}{1,0.4,0.2}

\usepackage{color}
\usepackage{aas_macros}
\usepackage{amsmath}
\newcommand{\erg}{\mathrm{erg}}
\newcommand{\s}{\mathrm{s}}

\newcommand{\K}{\mathrm{K}}
\newcommand{\lbox}{L_\mathrm{box}}
\newcommand{\bh}{\mathrm{BH}}
\newcommand{\yr}{\mathrm{yr}}
\newcommand{\mfeed}{\dot{M}_\mathrm{feed}}
\newcommand{\mfeedn}{\dot{M}_\mathrm{feed,0}}
\newcommand{\sch}{\mathrm{sch}}
\newcommand{\AU}{\mathrm{AU}}
\newcommand{\kb}{k}
\newcommand{\eV}{\mathrm{eV}}
\newcommand{\keV}{\mathrm{keV}}
\newcommand{\bandkeV}[2]{[{#1}\ \keV,\ {#2}\ \keV]}
\newcommand{\cm}{\mathrm{cm}}
\newcommand{\km}{\mathrm{km}}
\newcommand{\hh}{\mathrm{H}}
\newcommand{\Hz}{\mathrm{Hz}}
\newcommand{\sr}{\mathrm{sr}}
\newcommand{\g}{\mathrm{g}}
\newcommand{\e}{\mathrm{e}}
\newcommand{\T}{\mathrm{T}}
\renewcommand{\d}{\mathrm{d}}
\newcommand{\jexp}{\langle j \rangle}
\newcommand{\ff}{\mathrm{ff}}

\newcommand{\dyn}{\mathrm{dyn}}
\newcommand{\edd}{\mathrm{Edd}}
\newcommand{\cool}{\mathrm{cool}}
\newcommand{\comp}{\mathrm{IC}}

\newcommand{\feed}{\mathrm{feed}}
\newcommand{\tran}{\mathrm{tran}}
\newcommand{\grav}{\mathrm{grav}}
\newcommand{\vir}{\mathrm{vir}}
\newcommand{\kn}{\mathrm{KN}}
\newcommand{\ph}{\mathrm{ph}}
\newcommand{\uv}{\mathrm{UV}}
\newcommand{\msun}{M_\odot}
\newcommand{\mdot}{\dot{M}}
\newcommand{\aox}{\alpha_\mathrm{ox}}

\newcommand{\vvec}{\mathbf{v}}
\begin{document}

\shorttitle{AGN X-ray emission}

\title{Inflow Generated X-ray Corona Around Supermassive
  Black Holes and Unified Model for X-ray Emission}
\author{Lile Wang$^1$ and Renyue Cen$^2$}

\footnotetext[1]{Princeton University Observatory,
  Princeton, NJ 08544; lilew@astro.princeton.edu}

\footnotetext[2]{Princeton University Observatory,
  Princeton, NJ 08544; cen@astro.princeton.edu}

\begin{abstract}
  Three-dimensional hydrodynamic simulations, covering the
  spatial domain from hundreds of Schwarzschild radii to
  $2~$pc around the central supermassive black hole of mass
  $10^8\msun$, with detailed radiative cooling processes,
  are performed.  Generically found is the existence of a
  significant amount of shock heated, high temperature
  ($\ge 10^8\ \K$) coronal gas in the inner
  ($\le 10^4 r_\sch$) region.  It is shown that the
  composite bremsstrahlung emission spectrum due to coronal
  gas of various temperatures are in reasonable agreement
  with the overall ensemble spectrum of AGNs and hard X-ray
  background.  Taking into account inverse Compton
  processes, in the context of the simulation-produced
  coronal gas, our model can readily account for the wide
  variety of AGN spectral shape, which can now be understood
  physically.  The distinguishing feature of our model is
  that X-ray coronal gas is, for the first time, an integral
  part of the inflow gas and its observable characteristics
  are physically coupled to the concomitant inflow gas.  One
  natural prediction of our model is the anti-correlation
  between accretion disk luminosity and spectral hardness:
  as the luminosity of SMBH accretion disk decreases, the
  hard X-ray luminosity increases relative to the UV/optical
  luminosity.
\end{abstract}
\keywords{accretion, accretion disks --- galaxies: active
  --- galaxies: nuclei --- quasars: general --- X-rays:
  galaxies --- methods: numerical}

\section{Introduction}
\label{sec:introduction}

According to the commonly accepted paradigm, a luminous
Active Galactic Nucleus (AGN) has a supermassive black hole
at the center embedded in an equatorial accretion disc. The
hard non-thermal emission is generally thought to arise in a
hot gas above the accretion disc \citep[e.g.,][]{2000Elvis}.
% AGNs are the main contributor to the meta-galactic hard
% X-ray background \citep[e.g.,][]{2003GandhiFabian}.

Hot accretion flow based on the vertical one-zone
approximation is shown to possibly play a role, coming in a
variety of flavors, such as advection-dominated accretion
flow, convection-dominated accretion flow, and
advection-dominated inflow-outflow solution
\citep[e.g.,][]{1977Ichimaru, 1994Narayan, 1999Blandford,
  2000Igumenshchev}, some of which can reasonably fit the
observed X-ray spectra of low-luminosity AGNs
\citep[e.g.,][]{1998Narayan}.  The concurrent soft and hard
X-ray components observed in ordinary AGNs and the high
X-ray luminosity are not well modeled by the one-zone
hot-flow models.  A composite disk-corona structure may be
required.  The generally accepted model is that UV/optical
seed photons from the disc powered by accretion power are
Compton upscattered by hot thermal electrons in the corona.
Such a physical process leads to a power-law spectrum
extending to energies determined by the electron temperature
in the hot corona.  The power-law index is a function of the
plasma temperature and optical depth \citep[e.g.,][]
{1980Sunyaev, 1991Haardt, 1994Haardt, 1995Pietrini,
  2011Veledina}.  In addition, the primary X-ray radiation
in turn illuminates the disc and it is partly reflected
towards the observer's line of sight, often in fluorescence
from near-neutral (for iron) material of the disc, that has
been observed \citep[e.g.,][]{1989Fabian, 1995Tanaka,
  2003Reynolds}.
 
Since the spectrum depends on the optical depth and the
temperature of the hot gas, spectral shape can be reproduced
by some combinations of these parameters.  This successful
model has a major missing element.  That is, we do not know
what heats the corona plasma to temperatures as high as
$10^9\ \K$.  Heating via magnetic field reconnection
\citep[e.g.,][]{1998DiMatteo, 2000Miller, 2002Liu, 2003Liu,
2011Ohsuga} may be a relevant process, perhaps operating
in a fashion similar to solar corona.

Here we explore the dynamics of gas inflows using
three-dimensional hydrodynamic simulations to attempt to
construct an alternative model.  Our approach is to survey
the gas feeding conditions at large scales and examine how
they affect the distribution of shock heated gas powered by
gravitational energy of inflow gas.  Our simulations show
that hot gas of temperatures often exceeding $10^9\ \K$,
resembling the purported corona, can be naturally produced
in the inner regions of $\lesssim 3\times 10^3r_\sch$, with
hotter gas being closer to the black hole.  The composite
bremsstrahlung emission alone from the hot gas with a range
of temperatures may be able to account for a significant
fraction, if not most, of the X-ray emission observed in
high (optical/UV) luminosity AGNs, and of the hard X-ray
background.  We combine the simulation results with a
treatment of the inverse Compton process to show that the
observed anti-correlation between the optical/UV luminosity
and the hardness of the optical-X-ray spectral index is
naturally explained.  The physical origin is of
self-regulation in nature.  At high optical-UV luminosity of
the AGN, the high inverse Compton cooling rate renders the
inner region devoid of significant hot coronal gas.  Thus,
the inverse Compton optical depth and overall temperature of
the coronal gas are low.  The resulting, relatively weak
X-ray emission in this case is primarily contributed by
bremsstrahlung process.  At low optical-UV luminosity, the
low inverse Compton cooling rate results in a more centrally
peaked and overall higher inverse Compton optical and
coronal gas temperature, which in turn yield harder and
comparatively higher inverse Compton emission that is now
dominant.  We show that our model explains remarkably well
the observed range of optical-X-ray spectral index for
Eddington ratios ranging from $10^{-3}$ to $0.3$.  This
paper is structured as follows.  Section
\ref{sec:method-model} describes the setup of our
simulations and methods of analysis.  Results are presented
in Section \ref{sec:results}, followed by conclusions in
Section \ref{sec:summary}.

\section{Computational Methods}
\label{sec:method-model}

\subsection{Hydrodynamic Code}
\label{sec:algor-hydro}

We perform simulations of gas flows around supermassive
black holes with the widely used adaptive mesh refinement
(AMR) Eulerian hydro code, Enzo \citep[][]{2014Bryan}. Our
simulation domain covers a three dimensional cubic box with
side length $\lbox = 4\ \pc$, in which Cartesian coordinates
are initially established.  The root grid (i.e., the base
level) is $128^3$, upon which additional refinement levels
may be applied.  We find, upon inspecting several running
examples, that the ``effective'' regions in which refinement
should be applied turns out to be relatively invariant.
Thus, we initially lay down a set of statically refined grid
with five levels, with a refinement factor of two for each
level, detailed in Table \ref{tab:refinement-layout}.  We
also run one simulation run with six levels of refinement
to test numerical convergence.

%%%%%%%%%%%%%%%%%%%%%%%%%%%%%%%%%%%%%%%%
\begin{table}% [centering]
  \centering
  \caption{Geometry of refinement regions. Those are
    rectangular regions in $x\times y\times z$ in units of
    $\lbox$, centered at the box center.} 
  \vspace{0.5cm}
  \begin{tabular}{cccc}
    \hline
    Region No. & Level & Shape & Dimension (in $\lbox$) \\
    \hline
    1 & 1 & rectangular & $0.5\times 0.5 \times 0.05$ \\
    2 & 2 & rectangular & $0.16 \times 0.16 \times 0.02$ \\
    3 & 3 & rectangular & $0.08 \times 0.08 \times 0.01$ \\
    4 & 4 & rectangular & $0.04 \times 0.04 \times 0.01$ \\
    5 & 5 & rectangular & $0.04 \times 0.04 \times 0.01$ \\
    6 & 6 & rectangular & $0.04 \times 0.04 \times 0.01$ \\
    \hline
  \end{tabular}
  \label{tab:refinement-layout}
\end{table}
%%%%%%%%%%%%%%%%%%%%%%%%%%%%%%%%%%%%%%%%

We solve the hydrodynamics equations with Newtonian gravity.
\begin{equation}
  \label{eq:hydro}
  \begin{split}
    \dfrac{\partial \rho}{\partial t} + \nabla \cdot ( \rho
    \vvec ) = & 0\ ,\\
    \dfrac{\partial \vvec}{\partial t} + ( \vvec \cdot
    \nabla ) \vvec + \dfrac{1}{\rho} \nabla p + \nabla \Phi
    = & 0\ ,\\ 
    \dfrac{\partial e}{\partial t} + (\vvec \cdot \nabla) e
    + \dfrac{p}{\rho} \nabla \cdot \vvec = &\Gamma-\Lambda \ ,\\
    \nabla^2 \Phi = & 4\pi G \rho \ ,
  \end{split}
\end{equation}
where $\rho$, $p$, $e$, $\vvec$ and $\Phi$, indicate
  mass density, thermal pressure, specific internal energy
  per unit mass,velocity vector and gravitational potential,
  respectively; $G$ is the gravitational constant.
There are several hydro modules implemented in ENZO.
  We use the most robust ZEUS hydro solver, which
  incorporates artificial viscosity to capture shocks,
  producing correct Rankine-Hugoniot conditions across shock
  fronts \citep[see also][]{1992StoneZeus}.
We note here that for the energy equation in Equation
  \ref{eq:hydro}, $\Lambda$ and $\Gamma$ are the total of
  radiative cooling and heating rates, respectively.

The radiative cooling term $\Lambda$ includes all cooling
  processes from a primordial gas and metal cooling
  processes assuming solar metallicity. The heating term
  mainly consists of photoionization heating of all species
  in the gas and Compton heating.  This heating term is not
  treated in this work (i.e.  $\Gamma = 0$) due to lack of
  self-consistent calculation of radiative transfer, which
  thus has enhanced cooling of gas the simulations.  Thus,
  our findings of very hot gas in the central region is
  likely on the conservative side.
  
% These two terms include all radiative processes for a
%   primordial gas, including free-free cooling
%   \citep{1997Abel} and for metal cooling processes
%   \citep{2008MNRAS.385.1443S}. 

\subsection{Setup}

\label{sec:init-bound-cond}

In each model, a supermassive black hole of mass
$M_\bh = 10^8\msun$ is placed at the center of the box, 
with a gravitational acceleration
\begin{equation}
  \label{eq:grav-accel}
  \mathbf{g} = - \dfrac{G M_\bh \hat{\mathbf{r}}}{r_0^2}\ ,\quad
  r_0 = \begin{cases}
    &r\ ,\ r > a\ ;\\
    &a\ ,\ r \leq a\ ;\\
  \end{cases}
\end{equation}
where $a$ is the core radius, which may be varied; $r$ is
the distance from the black hole, and $\hat{\mathbf{r}}$ is
the unit vector pointing at the cell in question from the
SMBH. This softening radius, set to be significantly larger
than the hydro resolution, allows us to gauge how
gravitational heating affects the gas density and
temperature structure and also (easily) avoid numerical
singularity near the black hole.

An accretion disk of mass $\sim 10^7\msun$ surrounds the
SMBH.  Rather than placing the disk by hand, it is built by
feeding gas with the same pattern (angular momentum
distribution) but with a higher feeding rate for about
$\sim 2\times 10^4\ \yr$, after which we turn off gas
feeding at the boundary so that the disk structure can relax
for $\sim 3\times 10^4\ \yr$.  This point, at the end of the
disk relaxation, marks time $t=0$ for each of the
simulations.  We do not include self-gravity of the gas disk
and other gas.  Realistic gas feeding rate starts at $t=0$.
We use data outputs at $t>10^4\ \yr$ only, to allow for the
gas feeding pattern to reach a statistically steady state,
noting that the free fall time at the boundary of the box is
$5.7\times 10^3\ \yr$. Data outputs in
$t=10^4-2\times 10^4\ \yr$ are used for all subsequent
analysis, with a frequency of one output every $10\ \yr$.

At the outer boundaries, gas inflows are fed through a
fraction of random grid cells.  The rest of the cells are
marked as ``free'' (or ``Dirichlet'') boundary.  Boundary
feedings has specific fixed value of total mass inflow rate,
denoted as $\mfeed$.  Gas feeding is through 10 percent of
the randomly chosen ``pixels'' of the feeding faces.  We do
not feed gas from the top and bottom faces of the simulation
box.  Among the four side faces, we vary the number of faces
for feeding to assess the effects of feeding (a)symmetries.
Thus, the feeding rate at a given pixel on the side faces is
%%%%%%%%%%%%%%%%%%%%%%%%%%%%%%%%%%%%%%%
\begin{equation}
  \label{eq:feed-space-distrib}
  \mdot_\mathrm{feed,\ pixel} = \dfrac{\mfeed}{0.1 N_\feed}
  \left( \dfrac{\delta l}{\lbox}\right)^2 \Theta( 0.1 - X )\ ,
\end{equation}
where $\delta l = L_\mathrm{box}/128$ is the pixel size at
the boundary, $N_\feed=1,\ 2,\ 4$ is the number of side
faces that is fed through, $X$ is a random variable obeying
the uniform distribution $U(0,1)$.  $\Theta(x)$ is the
Heaviside step function, where $\Theta(x) = 1$ for all $x>0$
and $\Theta(x) = 0$ for all $x<0$.  This randomness is
introduced to emulate stochastic feeding.  However, in our
simulations, we find that the result is little changed, in
terms of central coronal density and temperature profile and
hence radiative characteristics, if we change the fraction
through which the gas is fed, or totally remove this random
feeding scheme and adopt a steady feeding across chosen
boundary cells.

% which verifies the robustness of our proposed mechanism.

%%%%%%%%%%%%%%%%%%%%%%%%%%%%%%%%%%%%%%%

The inflow gas fed through the outer boundary obeys the
following specific angular momentum distribution:
%%%%%%%%%%%%%%%
\begin{equation}
  \label{eq:feed-j-distrib}
  \dfrac{\d \mfeed(j)} {\d j} = 
  \begin{cases}
    \dfrac{\mfeed}{2\jexp}\ ,\quad & 0 < j < 2\jexp\ ;\\
    0\ ,\quad & \mathrm{elsewhere}\ ,
  \end{cases}
\end{equation}
%%%%%%%%%%%%%%%
with the expectation value $\jexp$ that may vary for
different models.  Note that such an angular momentum
distribution would correspond to the classic
\citet{1963Mestel} disk, if allowed to settle on its own
self-gravity.  Such a distribution is an excellent
approximation to results from realistic, high resolution
($\le 0.1\ \pc$) simulations \citep{2010Hopkins,
  2011Hopkins}.

%%%%%%%%%%%%%%%%%%%%%%%%%%%%%%%%%%%%%%%%
\begin{table*}% [centering]
  \centering
  \caption{Parameters of simulations.}
\vspace{0.5cm}
  \begin{tabular}{ccccccc}
    \hline
    Model No. & $\mfeed/\mfeedn$ & $\jexp/j_0$ & $a/a_0$ & Feeding
    face & Static Refinement Levels &\\
    \hline
    0  & 1     & 1     & 1    & 1 & 5 \\
    1  & 1     & 0.1   & 1    & 1 & 5 \\
    2  & 1     & 0.01  & 1    & 1 & 5 \\
    3  & 1     & 0.001 & 1    & 1 & 5 \\
    4  & 0.1   & 0.001 & 1    & 1 & 5 \\
    5  & 0.01  & 0.001 & 1    & 1 & 5 \\
    6  & 0.005 & 0.001 & 1    & 1 & 5 \\
    7  & 0.002 & 0.001 & 1    & 1 & 5 \\
    8  & 0.001 & 0.001 & 1    & 1 & 5 \\     
    9  & 0.01  & 0.1   & 1    & 1 & 5 \\
    10 & 0.01  & 0.01  & 1    & 1 & 5 \\       
    11 & 10    & 1     & 1    & 2 & 5 \\
    12 & 3     & 1     & 1    & 2 & 5 \\
    13 & 1     & 1     & 1    & 2 & 5 \\
    14 & 0.3   & 1     & 1    & 2 & 5 \\
    15 & 0.1   & 1     & 1    & 2 & 5 \\
    16 & 1     & 0.5   & 1    & 2 & 5 \\
    17 & 1     & 2     & 1    & 2 & 5 \\
    18 & 1     & 5     & 1    & 2 & 5 \\    
    19 & 1     & 1     & 0.5  & 2 & 5 \\
    20 & 1     & 1     & 0.25 & 2 & 5 \\
    21 & 1     & 1     & 1    & 4 & 5 \\
    22 & 1     & 1     & 1    & 2 & 6 \\    
    \hline
  \end{tabular}
  \label{tab:simulation-series}
\end{table*}
%%%%%%%%%%%%%%%%%%%%%%%%%%%%%%%%%%%%%%%%

The set of simulations performed is listed in Table
\ref{tab:simulation-series}.  Run ``0'' is our fiducial run
with $\mfeedn \simeq 8.1 \msun/\yr$, the expectation value
of the angular momentum distribution $j_0$ where the value
of $j_0$ is such that gas with $j_0$ is rotationally
supported when reaching $r\sim 0.4\ \pc$, gravitational
softening length $a=a_0\equiv1250 r_\sch$ in Equation
\eqref{eq:grav-accel} where $r_\sch$ is the Schwartzchild
radius of the SMBH at the center ($r_\sch = 1.97\ \AU$).
The 22 additional simulations are performed by varying one
of the five parameter at a time.  Specifically, we vary the
feeding angular momentum for 8 additional runs (Runs 1, 2, 3
at high feeding rate $\mfeed/\mfeedn = 1$ with
$\jexp/j_0 = 0.1,\ 0.01,\ 0.001$; Runs 9, 10 at low feeding
rate $\mfeed/\mfeedn = 0.01$ with $\jexp/j_0 = 0.1,\ 0.01$;
Runs 16, 17, 18 at high feeding rate $\mfeed/\mfeedn = 1$
with two feeding faces and $\jexp/j_0 = 0.5,\ 2,\ 5$), the
feeding rate for 10 additional runs (Runs 4, 5, 6, 7, 8 at
low angular momentum $\jexp/j_0 = 0.001$ with
$\mfeed/\mfeedn = 0.1,\ 0.01,\ 0.005,\ 0.002,\ 0.001$; Runs
11, 12, 13, 14, 15 at high angular momentum $\jexp/j_0 = 1$
with feeding rate $\mfeed/\mfeedn = 10,\ 3,\ 1,\ 0.3,\ 0.1$
and two feeding faces), the gravitational softening radius
for two additional runs (Runs 19, 20 with
$a/a_0 = 0.5,\ 0.25$) , the number of feeding faces for one
additional run (Run 21) and finally an additional run (Run
21) to vary the resolution.

\subsection{Post-Simulation Analysis}
\label{sec:analysis-method}

The X-ray emissivity by bremsstrahlung process (denoted by
subscript ``ff'') is \citep[e.g.][]{2011Draine}
\begin{equation}
  \label{eq:jff-emission-spectra}
  \begin{split}
    j_\ff & = 5.44\times 10^{-41} \erg\ \cm^{-3}\ \s^{-1}\
    \Hz^{-1}\ \sr^{-1}
    \\
    & \times g_\ff T_4^{-1/2} \left[\dfrac{n_in_e}{(1\ \g\
        \cm^{-3})^2}\right] Z_i^2 \e^{-h\nu/\kb T}\ ,
  \end{split}
\end{equation}
where $n_i$ and $n_e$ are the number density of ionized
particles, $T_4=(T/10^4\ \K)$ where $T$ is the plasma
temperature, $Z_i$ is the effective charge number of the
ions, and $g_\ff$ is the Gaunt factor,
\begin{equation}
  \label{eq:gaunt-factor}
  g_\ff \simeq \ln \left\{ \e + \exp\left[5.96 -
      \dfrac{\sqrt{3}}{\pi} \ln( Z \nu_9 T_4^{-3/2} )\right]
  \right\}\ ,
\end{equation}
in which $\nu_9=(\nu/10^9\ \Hz)$.  Given the simulation
output data, Equations \eqref{eq:jff-emission-spectra} and
\eqref{eq:gaunt-factor} will be applied to calculate the
spectra of free-free emission.  We note here that the
electron-electron bremsstrahlung is safely ignored, since
this only under-estimates the bremsstrahlung emission by a
few per cent in the $\bandkeV{10}{100}$ regime, at electron
temperature $T_e \sim 10^9\ \K$ \citep[see
also][]{1975Haug}.  We obtain total emission in different
energy bands by integrating Equation
\eqref{eq:jff-emission-spectra} through specific energy
ranges.
  
The disk is distinguished from coronal region.  We
procedurally exclude the disk by weighting the the computed
quantity in question by a factor $\e^{-\tau /2}$, where
$\tau$ is the optical depth of the ``pixel''.  Using a
different weighting factor $\e^{-\tau}$ does not produce
materially different results.  For example, we find that the
difference in total X-ray emission differs by about 3 per
cent between the two weighting schemes.  We also tested an
additional criterion, making use of the fact that the disk
gas usually obey $v_t > v_r$ (i.e., the tangential velocity
being greater than the radial velocity).  This additional
criterion, together with the optical depth one, gives a
composite weighting factor $\Theta(v_r-v_t)\e^{-\tau/2}$,
where $\Theta(x)$ is the Heaviside theta function with
argument $x$.  The overall results using this composite
weighting factor are very similar to that using
$e^{-\tau/2}$ factor alone.

\section{Results}
\label{sec:results}

\subsection{Existence of Generic Gravitational Shock Heated
  Coronal Gas}
\label{sec:corona}

\begin{figure}[!h]
  \centering
  \includegraphics[width=6.3in, keepaspectratio]
  {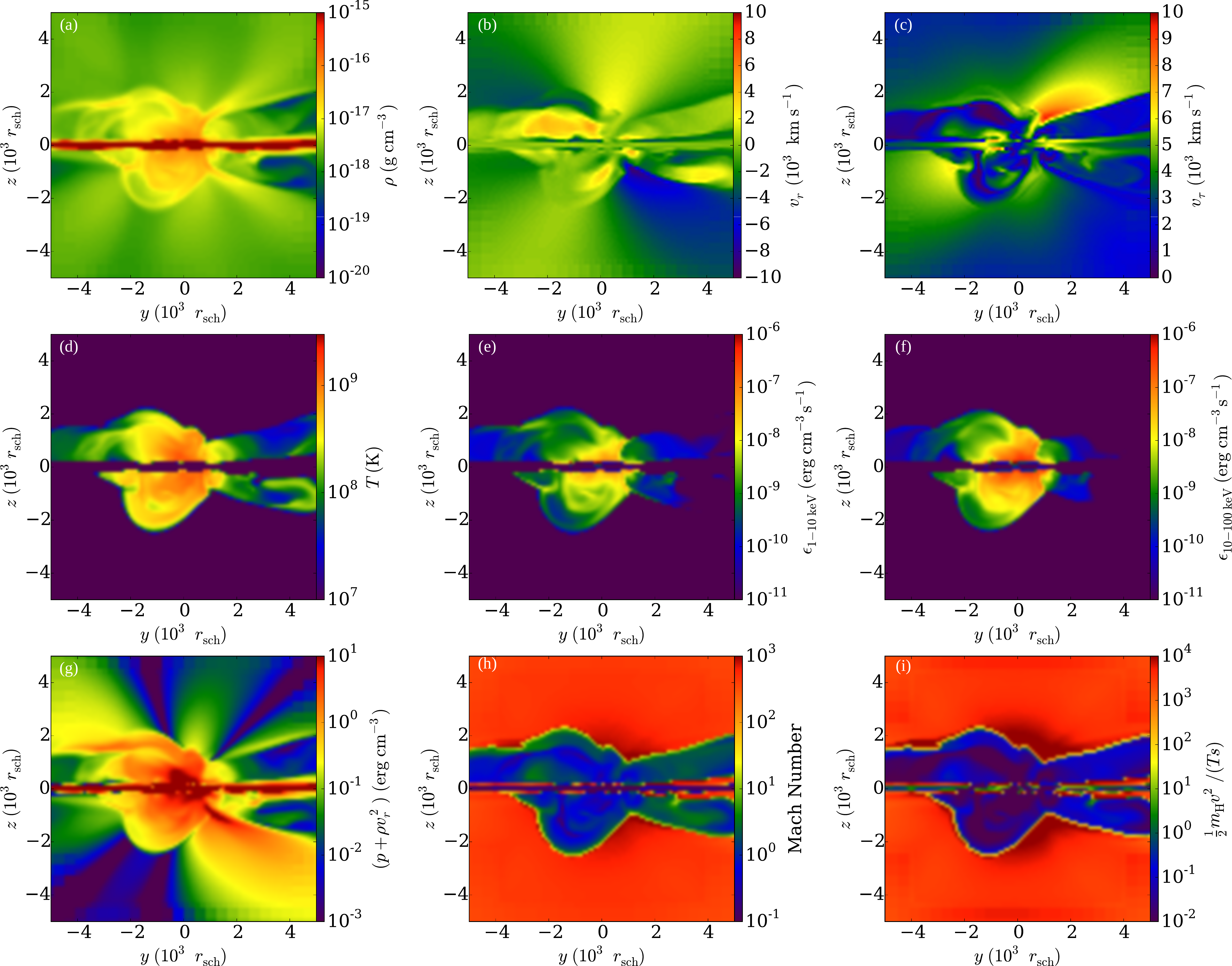}
  \caption{Shows the gas density (in $\g\ \cm^{-3}$, panel
    a), radial velocity ($v_r$ in $10^3\ \km\ \s^{-1}$,
    panel b), magnitude of tangential velocity ($v_\tau$ in
    $10^3\ \km\ \s^{-1}$, panel c), temperature (in Kelvin,
    panel d), integrated X-ray emissivity (in
    $\erg\ \cm^{-3}\ \s^{-1}$) in the $\bandkeV{1}{10}$
    band (panel e) and in the $\bandkeV{10}{100}$ band (panel
    f), total pressure (the sum of gas pressure $p$ and radial ram
    pressure $\rho v_r^2$, in $\erg\ \cm^{-3}$, panel
    g), Mach number (dimensionless, panel h), and ratio
    between single particle bulk motion kinetic energy
    ($m_\mathrm{H} v^2/2$) and $Ts$ (temperature times
    single particle Sackur-Tetrode entropy) (dimensionless,
    panel i), for a slice of size $10^4r_{\sch}$ cutting
    through the center and perpendicular to the disk, from
    the fiducial simulation (Run 0 in Table
    \ref{tab:simulation-series}).  }
  \label{fig:no_rad_slice}
\end{figure}
  
Figure \ref{fig:no_rad_slice} shows the gas density (in
$\g\ \cm^{-3}$, panel a), radial velocity (panel b),
tangential velocity magnitude (panel c), temperature (panel
d), X-ray emissivity in the $\bandkeV{1}{10}$ band (panel e)
and in the $\bandkeV{10}{100}$ band (panel
f), %(in $\erg\ \sˆ{-1}\ \cmˆ{-3}$)
total pressure (thermal pressure plus ram pressure, panel
g), Mach number (panel h), and ratio of kinetic and entropic
energy (calculated with Sackur-Tetrode entropy, panel i),
for a slice with one-cell thickness of size
$10^4r_{\sch}$ cutting through the center and perpendicular
to the disk from the fiducial simulation.
% We specially note here that the ``dynamic range'' of the
% color map is {\it not} the dynamic range of our
% simulation: narrower color ranges are adopted to present
% ranges of variation in which we are most interested.
It is noted that there is a cold ($\lesssim 10^4\ \K$, below
the range of the color map) disk at the equatorial plane due
to its high density and short cooling time.  Since we do not
properly treat the viscous energy dissipation in the disk as
well as the radiative losses, the temperature of the disk is
not correct.  Nevertheless, our calculation of X-ray
emission in the corona is not contingent upon a realistic
treatment of the dense disk.  It is possible that, due to
the disk colder than it should be, some coronal gas will be
more prone to sticking/accreting to the disk due to an
under-estimate of vertical pressure gradient near the disk.
Thus, it is likely that we have somewhat under-estimated the
hot coronal gas due to this mis-treatment of accretion disk.
The contrasting temperature, density and kinematics allow us
to easily separate the corona material from the disk.

The coronal regions, above and below the disk, are easily
distinguished from the disk by their high temperatures
$10^{8-9}\ \K$ and lower density.  The coronal regions of hot
gas are seen to occupy the central regions of size a few
times $10^3r_\sch$.  While we will show that the dynamics of
the coronal region depends on the boundary conditions
(feeding rate, angular momentum distribution, feeding
pattern, etc), our simulation box is more than ten times
larger than the coronal regions of interest and its size is
thus large enough to not impact the properties of the
coronal regions.  All our results are based on time
snapshots after $\gtrsim 10^4\ \yr$, compared to the free
fall time of $\sim 800\ \yr$ at $2\times 10^4r_\sch$, the
coronal regions with hard X-ray emission is fully relaxed
and basically in a statistically steady state.

The only heating mechanisms that are allowed here in the
simulation are shocks and adiabatic compression. Boundaries
of heated regions are seen to be quite sharp, suggesting
that the main heating mechanism is shock heating.  In order
to further confirm this, we show in panels (g), (h) and (i)
in Figure \ref{fig:no_rad_slice} the quantities that are
pertinent to shocks. Panel (h) shows clearly a large jump in
Mach number across the shock heated ``surface''.  Shock
dissipation is also consistent with panel (i) showing a
drastic increase inward of the entropy across the shocked
surface. Note that compression heating conserves entropy.
The total pressure (thermal + ram pressure) (panel g)
demonstrates dynamically that compression does not play a
role, since the pressure increases inward, instead of
outward.  In general, there is a trend that higher
temperature gas is more centrally concentrated than lower
temperature gas.  These features are generic among all
simulations with varying feeding parameters at the boundary.

\begin{figure}
  \centering
\vspace{0.5cm}
  \includegraphics[width=8.5cm, keepaspectratio]
{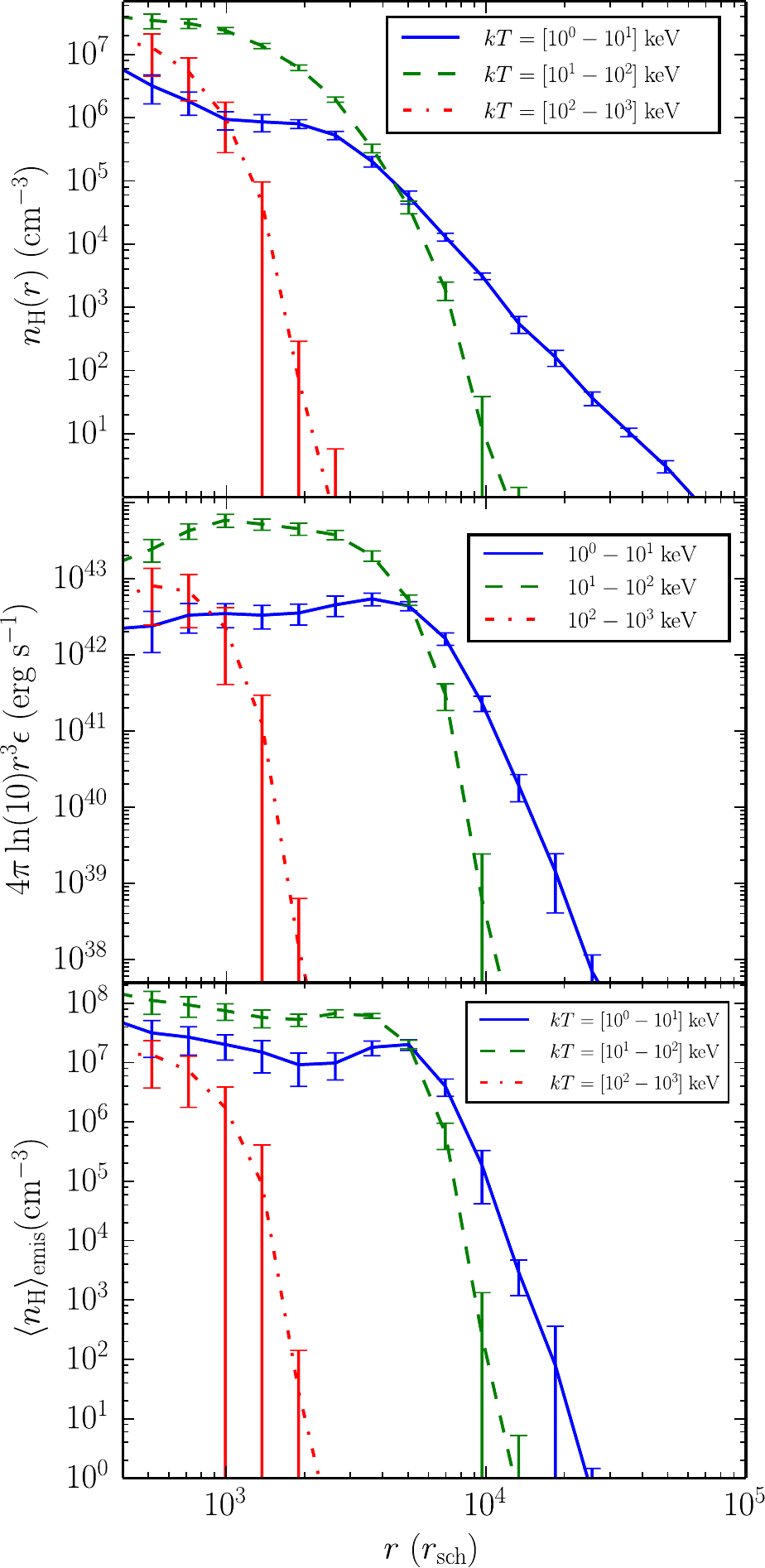} % File name changed already
  % {no_rad_f1_00_j1_00_radial_emissionweight.pdf}
\caption{ {\color{red} Top panel} shows the radial profile
  of density, averaged over all $4\pi$ solid angle, in three
  X-ray energy bands: $\bandkeV{1}{10}$ (blue solid),
  $\bandkeV{10}{10^2}$ (blue dashed), $\bandkeV{10^2}{10^3}$
  (red dot-dashed).  {\color{red} Middle panel} shows the
  radial distributions of luminosity, emitted by
  bremsstrahlung processes in the corona. Note that we have
  multiplied the emissivity $\epsilon$ by $4\pi \ln(10)r^3$
  so that the area below each line is the luminosity.
  {\color{red} Bottom panel} shows the radial profile of
  emissivity-weighted hydrogen density for the three
  respective bands.  We use 100 snapshots over a time period
  of $10^3\ \yr$ to compute the distribution, with the
  errorbars shown indicating the inter-quartiles.  }
  \label{fig:no_rad_fiducial_radial}
\end{figure}

Figure \ref{fig:no_rad_fiducial_radial} shows the radial
distribution profiles of density (top panel), bremsstrahlung
emissivity (middle panel) and emissivity-weighted hydrogen
density (bottom panel) of the coronal gas in three X-ray
bands. Note that we exclude the disk when making these
plots. In all panels, we confirm the visual evidence seen in
Figure \ref{fig:no_rad_slice} that the hard X-ray emission
is centrally concentrated.  For both $\bandkeV{1}{10}$ and
$\bandkeV{10}{10^2}$ bands, the X-ray emission mainly come
from the central $10^{4}r_\sch$, while for the
$\bandkeV{10^2}{10^3}$ band the central $10^{3}r_\sch$
dominates.
% Given the size of our simulation box, it is natural to
% expect that such hard X-ray emission is modeled without
% being limited by the finite size of our solution domain.
Significant fluctuations in X-ray emission on the time scale
of (at least as short as) $10\ \yr$ is seen, as indicated by
the errorbars, suggesting that the X-ray emitting regions
are not static, with higher temperature gas exhibiting
larger amplitude temporal fluctuations than lower
temperature gas. To place things in a more quantitative
context, the mean bremsstrahlung X-ray luminosity is
$1.65\times 10^{43}\ \erg\ \s^{-1}$ for $E>1\ \keV$, and
$1.17\times 10^{43}\ \erg\ \s^{-1}$ for $E>10\ \keV$,
compared to the Eddington luminosity of
$L_\edd(10^8\msun) \simeq 1.3\times10^{46}\erg\ \s^{-1}$ for
a $10^8\msun$ black hole.

It is useful to have a basic understanding of the gas
temperature.  If one equates the thermal energy ($3\kb T/2$)
of a particle around the SMBH to the free-fall energy at
radius $r$, one finds that,
\begin{equation}
  \label{eq:Tvir}
  T_\vir=6.3 \times 10^8\ \K \left(\dfrac{r} {3\times
      10^3r_\sch} \right)^{-1}\ .
\end{equation}
It is easy to see that the gas temperature obtained from the
simulation is in good accordance with this estimate, after
infalling gas being shock heated. From panel (i) of Figure
\ref{fig:no_rad_slice}, we clearly observe that the gas
preserve its bulk motion kinetic energy above the shock
surface, at which kinetic energy is thermalized. Needless to
say, the ultimate energy source is gravitational. The amount
of feeding gas reaching various radii will depend on the
physical conditions and dynamics of the gas, with varying
amounts of gas at varying radii cooling out from the inflow
to join the disk. Therefore, the intensity of X-ray emission
out of this dynamic mechanism will likely be determined by
the amount of radial mass flux into the region of
$R\lesssim 3\times 10^3r_\sch$ if the gas cooling time is
shorter than the dynamical time, by the amount of hot gas
accumulated over the cooling time if the cooling time is
longer than the dynamical time.

Coulomb collisions are the main physical mechanism that
equilibrates electron and ion temperatures. The
equilibrating timescale at radius $r$ for a fully ionized
medium \citep[][]{1962Spitzer}, comprised of electrons,
protons, and He III, is
\begin{equation}
\begin{split}
  \label{eq:tei}
  t_\mathrm{ei} & \simeq 2\times 10^{-2}t_\dyn \left({r\over
      3\times 10^3r_\sch}\right)
  \left({T_e\over 10^8{\ \K}}\right) \\
  & \times \left({n_\hh\over 10^7\ \cm^{-3}}\right)^{-1}
  \left({\ln\Lambda\over 40}\right)^{-1}\ ,
\end{split}
\end{equation}
where 
\begin{equation}
\begin{split}
\label{eq:tdyn}
t_\dyn \simeq 10^3\ \s\ \left({M_\bh\over 10^8\msun}\right)
\left({r\over r_\sch}\right)^{3/2}\ ,
\end{split}
\end{equation}
is the free-fall time (viz. the dynamical time scale), $T_e$
the notional electron temperature, $n_\hh$ the hydrogen
number density, $\Lambda$ the Coulomb logarithm.  Given the
typical density $n_\hh\sim 10^6-10^7\ \cm^{-3}$, as is seen in
Figure \ref{fig:no_rad_fiducial_radial}, we conclude that
treating electrons and ions as a single temperature fluid is
a reasonable approximation.  This ensures that our
calculation of X-ray emissions, using the shock heating
temperature, from both the free-free process and inverse
Compton (see below) processes, is valid.

\subsection{Dependence of Coronal Gas on Boundary
  Feeding Conditions}
\label{sec:boundary}

\begin{figure*}
  \centering
\vspace{0.5cm}
  \includegraphics[width=4.8in, keepaspectratio]
  {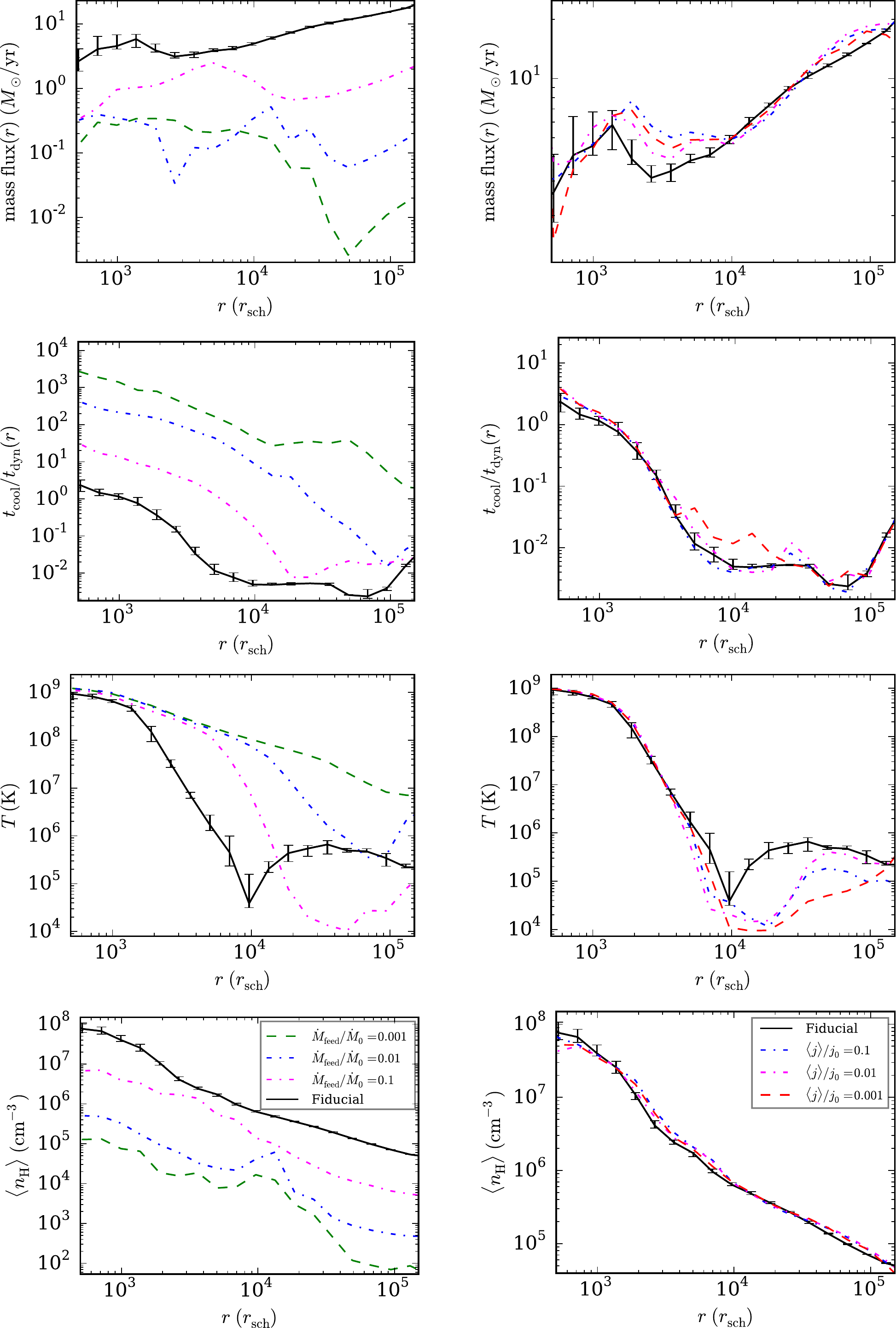}
  \caption{ {\color{red} The top row} shows the radial
    profiles of mass flux (integrated over $4\pi$ solid
    angle).  {\color{red} The second row} shows the
    bremsstrahlung cooling time normalized by the local
    dynamical time $t_\mathrm{dyn}$.  {\color{red} The third
      row} shows the gas temperature.  {\color{red} The
      bottom row} shows the gas number density
    $\langle n_\mathrm{H} \rangle$. {\color{red} The left
      column} shows comparisons among simulations with
    different feeding rate (Runs 0, 4, 5, 8), while
    {\color{red} the right column} varies the expectation
    value of specific angular momentum of feeding gas at the
    boundary of our simulation domain (Runs 0, 1, 2, 3). All
    statistics are calculated by averaging through $4\pi$
    solid angle and excluding the cold disk. Each curve is
    the result of statistics about a single simulation run,
    showing the median through 100 data dumps.  Error bars,
    presenting the 50 per cent frequency range, are only
    plotted for the ``fiducial'' simulation run.}
  \label{fig:no_rad_mass_flux_tcool}
\end{figure*}

We now show the impact of different boundary feeding
conditions on the concerned coronal gas.  We first examine
how the basic physical quantities in the X-ray emitting
regions depend on the gas feeding conditions at the
boundary, utilizing the set of simulations listed in Table
\ref{tab:simulation-series}. As indicated by simulation
results so far, the gas flowing inward is subject to both
heating and cooling processes, which, together with
dynamical interactions, such as angular momentum exchanges
among gas, determine the amount of mass flux into the inner
regions. Those in turn determine the X-ray emission at
various energy bands.  Figure
\ref{fig:no_rad_mass_flux_tcool} shows the radial mass flux
(first row; positive for inflow), the bremsstrahlung cooling
time in units of free fall time (second row), gas
temperature (third row), and gas number density (bottom
row). In all panels, the cold dense disk is excluded using
the criteria introduced in Section
\ref{sec:analysis-method}.  It is readily seen that the
amount of mass flux (top row) getting into the region of
several $10^3r_\sch$ is dependent strongly on the boundary
feeding rate (see the left column) and relatively weakly on
the initial specific angular momentum distribution (see the
right column).  The higher the gas feeding rate at the
boundary, the higher the amount of gas reaches the inner
regions.
% This explains the dependence of luminosity on gas feeding
% rate seen in top-left panel of Figure
% \ref{fig:no_rad_emis_spec_compare}.  The relatively weak
% dependence on the initial angular momentum of gas feeding
% at the boundary, is consistent with results seen in the
% bottom-left panel of Figure
% \ref{fig:no_rad_emis_spec_compare}.

What has happened to the gas that does not reach the inner
regions?  To answer this question, we examine the second row
of Figure \ref{fig:no_rad_mass_flux_tcool}, which show the
cooling time as a function of radius.  It is seen that, in
all cases, the ratio of cooling time to dynamical time,
$t_\cool/t_\dyn$, is always less than unity until the gas
has reached within about $10^{4}r_\sch$, the dominant X-ray
emitting region that we see in the simulations.  This
re-enforces the statement that the heating process is mainly
shock heating in stead of adiabatic compression, because the
adiabatic heating time scale is likely to be comparable to
the dynamical free-fall time scale.  It is instructive to
note that the cooling time of gas in the $r\sim 10^4r_\sch$
regions tends to be anti-correlated with the gas feeding
rate; i.e., the higher the gas feeding rate, the larger the
amount of gas that enters the inner regions, and hence the
shorter the cooling time.  The ``saturation'' of X-ray
luminosity with respect to feeding rate is attributed to
this, reflecting that higher cooling rate prevents gas
emissivity from going up linearly with feeding rate in part.
Throughout most of the radial range, mass outflow is
negligible in the runs with high feeding rate. However, due
to the very slow speed of cooling compared to dynamical
time, fluctuations of radial flow becomes important at small
radii, making it impossible to obtain accurate statistics
therein--especially for the runs with low feeding rate,
where the radial mass flux becomes unreliable at small
radii, because ``apparent'' inflow and outflow due to
velocity fluctuations become dominant.

Using Equation \ref{eq:Tvir} to define the hot coronal gas
temperature as a function of radius, we find that the gas
cooling time due to bremsstrahlung emission is,
\begin{equation}
\label{eq:tff}
\begin{split}
  t_\ff &= {3\kb T_\vir\over 2n_\hh\Lambda(T_\vir)} \\
  & \simeq 3.8\times 10^9\ \s\ \left({r\over 3\times
      10^3r_\sch}\right)^{-1/2}
  \left({n_\hh\over 10^6\ \cm^{-3}}\right)^{-1} \\
  & \simeq 17 \ t_\dyn \left({M_\bh \over
      10^8\msun}\right)^{-1} \left({r\over 3\times
      10^3r_\sch}\right)^{-2} \left({ n_\hh\over 10^6\
      \cm^{-3}}\right)^{-1}\ .
\end{split}
\end{equation}
Given the gas density seen in Figure
\ref{fig:no_rad_fiducial_radial}, we can now understand the
behavior of cooling time seen in the second row of Figure
\ref{fig:no_rad_mass_flux_tcool}.  Gas cooling time becomes
longer than the dynamical time only within a radius of
several times $10^3r_\sch$ at a density of
$n_\hh\sim 10^7\ \cm^{-3}$.  It is informative to notice
that the cooling time at $r\le 10^3r_\sch$ in all cases is
longer than the dynamic time, seen in the bottom panels of
Figure \ref{fig:no_rad_mass_flux_tcool}.  This indicates
that a significant amount of hot thermal gas of temperature
$\ge 10^9\ \K$ may be ``sustained'' for at least several
local dynamical times, even in the absence of continued
supply of gas (and hence energy).  On the other hand, the
cooling time at $r\sim 10^3-10^4r_\sch$ is shorter than the
dynamical time, suggesting that it is more ``difficult'' to
maintain gas at temperature $\sim 10^{8-9}\ \K$ without
continued supply of gas.  This cooling constraint suggests
that there is a maximum possible luminosity, if one sets
$t_\ff =t_\dyn$ at each radius, which in turn limits the gas
density at each radius. We confirm those arguments by
inspecting the third row in Figure
\ref{fig:no_rad_mass_flux_tcool}, in which the
``non-cooling'' range of radius (varies with each different
feeding condition) shows consistent slope (approximately
$T\propto r^{-1}$ in temperature plots.
% We will return to this point later in the discussion
% section.
Inclusion of inverse Compton processes alters gas
properties in the inner regions, which will be quantified
later.

\begin{figure*}
  \centering
\vspace{0.5cm}
  \includegraphics[width=6.3in, keepaspectratio]
  {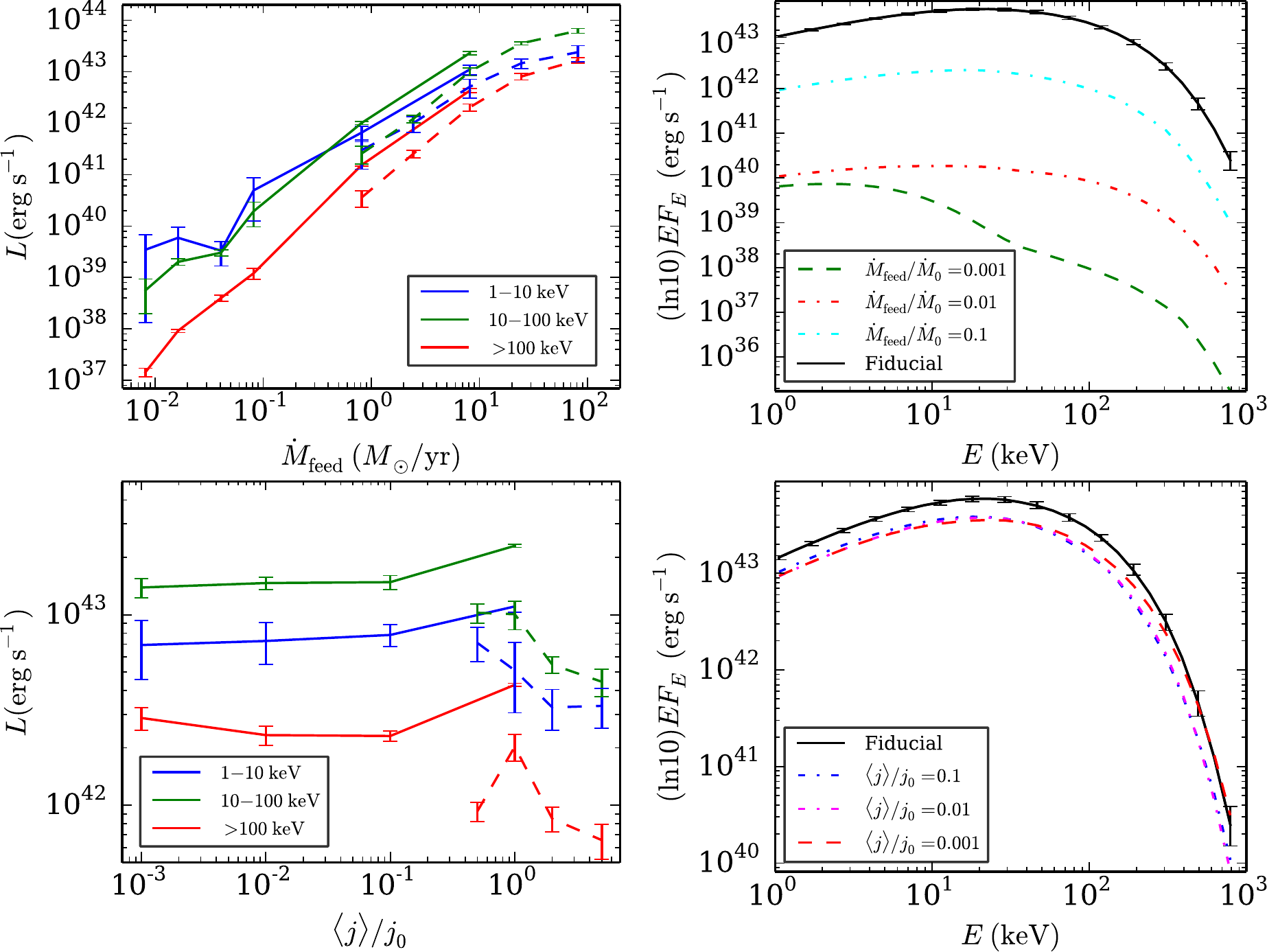}
  \caption{ {\color{red} Top-left panel} shows the X-ray
    luminosity in three bands as a function of gas feeding
    rate $\mfeed/\mfeedn$ for the one-face feeding with
    $\jexp/j_o=0.001$ (Runs 3,4,5,6) shown as solid curves
    and for the two-face feeding with $\jexp/j_o=1$ (Runs
    9,10,11,12,13) shown as dashed curves.  {\color{red}
      Bottom-left panel} shows the X-ray luminosity in three
    bands as a function of the expectation value of feeding
    specific angular momentum $\jexp/j_o$ with mass feeding
    rate fixed at $\mfeedn$ for the one-face feeding (Runs
    1,2,3,4) shown as solid curves and for the two-face
    feeding (Runs 11,14,15,16) shown as dashed curves, at
    the same $\mfeedn$.  {\color{red} Top-right panel and
      bottom-right panels} compare the emission spectra for
    the $\mfeed/\mfeedn$ series and the $\jexp/j_o$ series,
    respectively.  Each curve presents the median out of 100
    snapshots with the errorbars representing the
    interquartile, where shown.  For the spectra, we
    multiply each $F_E$ by $\ln(10)E$, so that integration
    of spectrum can be carried out along $E$ through the
    figures.  }
  \label{fig:no_rad_emis_spec_compare}
\end{figure*}

We now examine how the observable properties of the X-ray
emitting regions depend on the gas feeding conditions at the
boundary.  The top-left panel of Figure
\ref{fig:no_rad_emis_spec_compare} shows the X-ray
luminosity in three bands as a function of gas feeding rate
at a fixed $\jexp/j_o$ for the one-face feeding series
(solid curves) and the two-face feeding series (dashed
curves).  We see that the X-ray luminosity in all energy
bands increase with gas feeding rate monotonically.  Over
the entire range examined, the increase of X-ray luminosity
with increasing feeding rate is sublinear, indicating that
the amount of gas flux reaching the central regions
($\le 10^{4}r_\sch$) increases at a rate slower than the gas
feeding rate at the outer boundary.  In addition, there is
hint that the rate of increase of luminosity further
flattens significantly near the high end
($\dot M_{\rm feed}\ge 20\msun/$yr) of the feeding rate
range.  This ``saturation'' of X-ray luminosity with respect
to increasing feeding rate is attributed to the fact that
higher cooling rate prevents influx gas going up linearly
with feeding rate at the boundary, see in Figure
\ref{fig:no_rad_mass_flux_tcool}.

What is interesting is that, in the overlapping range of
$\dot M_{\rm feed}\sim 1-10\msun/$yr, the two series of runs
with very different $\jexp/j_o$ ($0.001$ for the solid
curves versus 1 for the dashed curves), seen in the top-left
panel of Figure \ref{fig:no_rad_emis_spec_compare}, differ
only by a factor of a few in X-ray luminosity at a given
$\dot M_{\rm feed}$.  The bottom-left panel of Figure
\ref{fig:no_rad_emis_spec_compare} shows X-ray luminosity as
a function of $\jexp/j_o$ and the dependence is a bit
complicated.  We see that, for the one-face feeding runs,
the X-ray luminosity {\it increases} mildly as the mean
specific angular momentum of the gas fed at the boundary,
$\jexp$, increases, with the same feeding rate $\mfeedn$ in
all cases.  One might have expected that, since a lower
$\jexp$ corresponds to a larger fraction of low angular
momentum gas at a given $\mfeedn$, the amount of gas
entering the central region would decreases with increasing
$\jexp$, which is opposite to the trend, albeit mild, seen.
This indicates that it is not the gas at the very low end of
the initial angular momentum distribution that goes into the
central regions, rather, the amount of gas actually going to
the central region is a result of interactions of gas of
varying initial angular momenta.  On the other hand, for the
two-face feeding runs, we do see that in most cases, the
X-ray luminosity {\it decrease} with increasing $\jexp$.
Overall, the findings suggest complexities in the gas
interactions that affect the eventual angular momentum
distribution at the low end, which presumably affects the
amount of gas eventually entering the central, shock heated
region powering the X-ray emission.

%%%%%%%%%%%%%%%%%%%%%%%%%%%%%%%%%%%%%
\begin{figure*}
  \centering
\vspace{0.5cm}
  \includegraphics[width=6.3in, keepaspectratio]
  {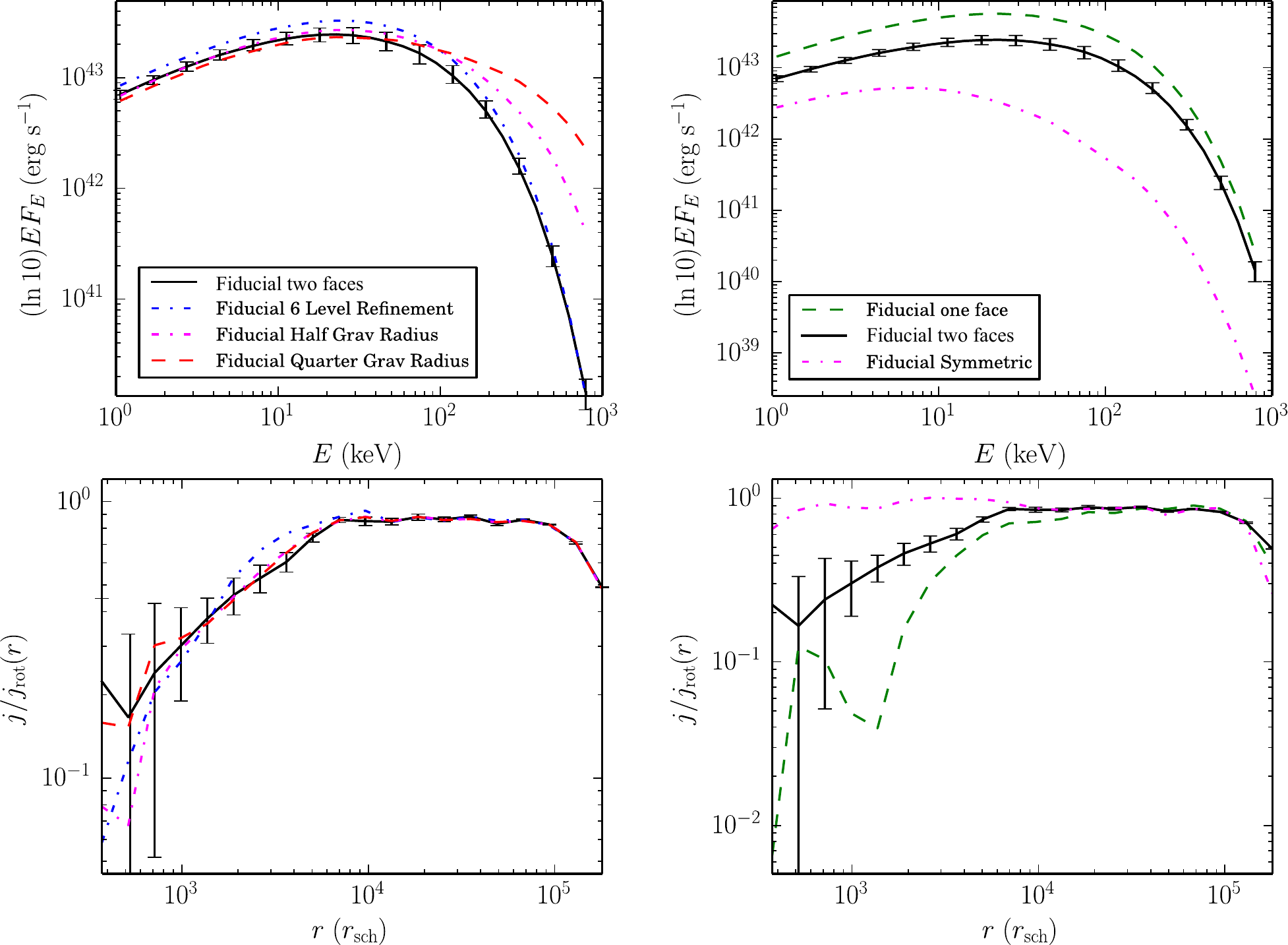}
  \caption{ {\color{red} Two left panels} show the radiation
    spectrum (upper panel) and specific angular momentum
    (lower panel), normalized by local circular rotation
    orbit specific angular momentum $j_\mathrm{rot}$), for
    the resolution series (Runs 0,17,18,20).  {\color{red}
      Two right panels} show the same results for feeding
    pattern series with $\mfeedn$ (Runs 0,11,19).  Each
    curve is based on 100 data dumps, with median and
    inter-quartiles shown only plotted for the ``fiducial''
    simulation run. }
  \label{fig:no_rad_consistency_check}
\end{figure*}
%%%%%%%%%%%%%%%%%%%%%%%%%%%%%%%%%%%%%

From the right panels of Figure
\ref{fig:no_rad_emis_spec_compare}.  it is seen that the
bremsstrahlung emission spectra are broadly peaked in the
energy band $\bandkeV{10^1}{10^2}$ in general for cases with
high enough feeding rate.  Above $10^2\ \keV$, the spectra
drop steeply. These spectral features are quite generic
for all simulation runs, indicating that the same shock
heating mechanism applies to hard X-ray emissions throughout
the parameter space represented by Figure
\ref{fig:no_rad_emis_spec_compare}.  For very low accretion
rate $\mfeed/\mfeedn=0.001$, a significant steepening of the
spectrum above $10\ \keV$ is seen, indicative of inability
of gas penetrating into the inner regions, which physically
is attributable to the increased pressure gradient in those
cases.  Those facts imply that feeding conditions are key
factors to the X-ray bremsstrahlung emission in the vicinity
of central black hole and at very low feeding rate, the
X-ray spectrum due to bremsstrahlung emission softens.

Here, we further examine how results depend on the feeding
patterns and also numerical resolution in Figure
\ref{fig:no_rad_consistency_check}.  In the upper-right
panel we see that feeding the box in a ``symmetric'' way
(i.e. via all four side faces) is actually giving
considerably lower amount of X-ray radiation than the other
cases, with identical total feeding rates in all cases.  To
provide some clue, the lower-right panel of Figure
\ref{fig:no_rad_consistency_check} shows the specific
angular momentum normalized by local circular rotation orbit
specific angular momentum.  We see that the the
``symmetric'' feeding pattern is able to sustain high
specific angular momentum - $j/j_\mathrm{rot}$ is always
close to unity - all the way down to the center.  In this
case, the gas that manages to inflow into the inner regions
is seen to be marginally on rotation support, which in turn
has posed a significant bottleneck on the amount of gas
inflowing; in other words, the amount of low angular
momentum gas is much smaller in this case compared to other
cases and in addition more of it is less prone to shocking
being on rotation support.  Thus, in this case, luminosity
is seen to be significantly lower and the spectrum
significantly softer, peaking in the band $\bandkeV{1}{10}$.
We see that the one-face feeding case produces a higher
luminosity than the two-face feeding case, which in turn is
higher than the ``symmetric'' feeding case.  If an infalling
cold stream or gas cloud gives rise to a high feeding rate
episode, one might argue that in this case the
``real-world'' situations would be better approximated by
``one face'' or a fraction of a face feeding.
% In the case of one-face feeding, it is seen that there is
% interaction between different gas parcels is relatively
% more reduced in the high feeding rate regime, so as to
% allow more gas on the $j \rightarrow 0 $ tail of the
% initial angular momentum distribution to reach central
% region directly.  To summarize, the majority fraction of
% the gas fed from the boundary will be rotationally
% supported at larger radii, and the amount of gas reaching
% the inner regions is the result of a complex interplay
% between angular momentum, shock heating and cooling.

In terms of the softening radius $a$, the spectrum is
hardened for a smaller $a$, as can be seen in Figure
\ref{fig:no_rad_consistency_check}.  This can be understood
as follows.  A smaller $a$ allows gas to continue to infall
towards smaller radius and heated to higher temperature.  We
expect that, with still higher resolution, the spectrum at
high energy end may drop off less steeply as shown in Figure
\ref{fig:no_rad_spec_observe} in the absence of other
physical processes that may be operating at small radii,
such as inverse Compton process, which we will discuss now.

\subsection{A Unified X-ray Emission Model Compared to Observations}
\label{sec:observations}

\label{sec:extrapolate}

In the simulations analyzed so far, we do not include
inverse Compton (IC) up-scattering of seed photons produced
by the accretion disk.  IC process would have two effects.
On one hand, the corona gas in the central region will be
cooled by the UV/optical photons emanating from the
accretion disk. On the other hand, the up-scattered
UV/optical photons will contribute to the X-ray emission
(and affect the UV/optical spectrum as well), which,
traditionally, is thought to be the primary X-ray emission
process of AGNs \citep[e.g.][]{1992walter, 1993haardt_b,
  1994Haardt}.  We now consider the effects of IC on gas
cooling and the resulting X-ray emission, jointly with the
free-free emission, in the context of the coronal gas
structure learned from the simulations.

\begin{figure}
  \centering
 \vspace{0.5cm}
  \includegraphics[width=6in, keepaspectratio]
  {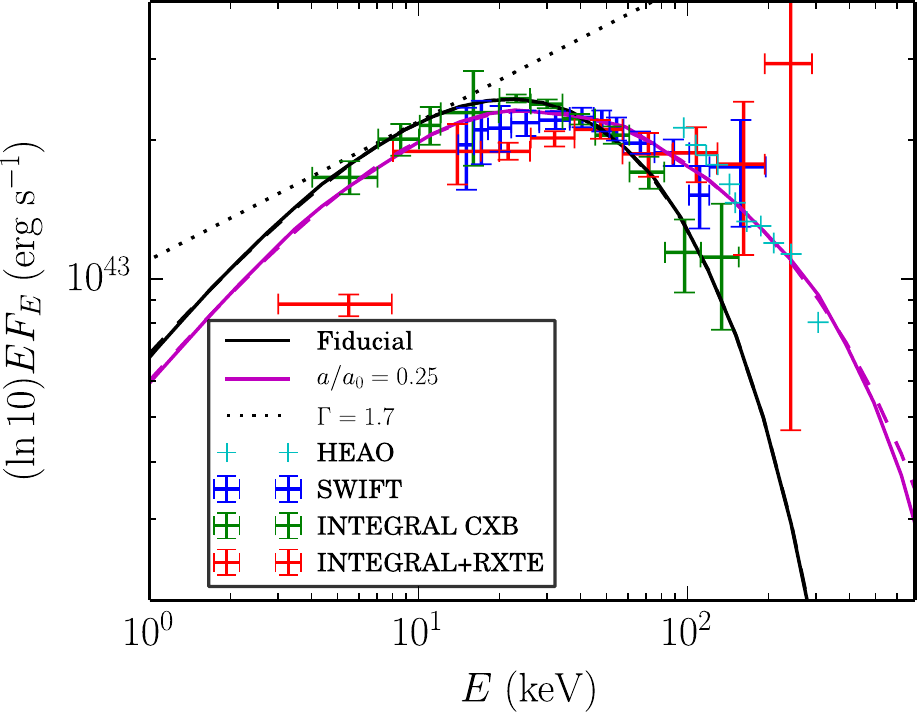}
  \caption{shows X-ray spectra of the simulation series with
    varying gravitational softening radius $a$ (the
    $a/a_0=0.5$ case lies in-between $a/a_0=0.25$ and
    $a/a_0=1$ and is omitted for display clarity).  Dotted
    line added to the figure indicates a pure power-law with
    photon index $\Gamma = 1.7$
    \citep[e.g.][]{2001Pappa}. Observation results are also
    overlaid onto the figure for comparison.  Unobscured
    composite AGN spectra from SWIFT observation
    \citep[see][]{2011Burlon} are indicated by blue
    errorbars.  Errorbars in green are the compilation of
    INTEGRAL X-ray background, while those in red are for
    composite AGN spectra observed by INTEGRAL and RXTE
    combined \citep{2008Sazonov}.  Data points coming from
    HEAO1-A4 MED shown by ``$+$'' in cyan without errorbars
    \citep{1997Kinzer}.}
  \label{fig:no_rad_spec_observe}
\end{figure}

Before we integrate IC process into our simulation results,
it is useful to make a direct comparison between X-ray
observations and simulation results without IC
contributions, shown in Figure
\ref{fig:no_rad_spec_observe}.  We clarify that the computed
spectra (curves in Figure \ref{fig:no_rad_spec_observe}) is
a composite bremsstrahlung emission from shock heated gas of
varying temperatures, primarily in the central $10^4r_\sch$
region, without any contribution from the gas disk.  We see
in Figure \ref{fig:no_rad_spec_observe} that, at the soft
energy end in the $\bandkeV{1}{5}$ range, the simulated
spectrum has too hard a slope.  Specifically, we find a
spectral index of $1.31$ in this energy range, which is to
be compared with the observed canonical value of
$\bandkeV{0.1}{0.3}$, ranging from $\sim 1.5$ to $\sim 2$,
\citep[e.g.][]{2001Pappa, 2002Gallagher,2012Falocco} from
stacked spectra in the $\bandkeV{1}{10}$ band.  We expect
that inclusion of metal line emission
\citep[e.g.][]{1995Cen, 2006Brocksopp} and contribution from
the accretion disk will likely soften the spectra, perhaps
bringing the results into better agreement with
observations.  Thus, since this energy range is not the
focus of our study, we are not worried about with this
apparent discrepancy at this time.

In the hard X-ray range of $\bandkeV{5}{300}$ that is our
focus, the simulated spectrum provides a good match to
observations.  Both the broad peak of the spectrum at
$\bandkeV{10}{30}$ and the hard X-ray spectral shape at
$\bandkeV{30}{300}$ are in good agreement with observations,
including the hard X-ray background that is dominated by
AGNs \citep{1997Kinzer, 2008Sazonov, 2011Burlon}.  Given our
completely un-fine-tuned simulation metrics, it is
non-trivial to achieve such a good agreement between generic
simulations and observations.  This is strongly indicative
that {\it the overall X-ray emission of all AGNs as an
  population can be largely accounted for by bremsstrahlung
  emission.}  Also implied by this observation is that the
overall hard X-ray background can be largely accounted from
contribution of high UV/optical luminosity AGNs.

To understand and account for the overall variety of X-ray
spectra of AGNs, inverse Compton processes may be
indispensable, as we show now.  In order to take IC
processes into account, it is necessary to extrapolate our
simulation results of density and temperature to smaller
radii, in part because of resolution limitations and in part
because of neglect of IC processes in the simulations.
Based on Figures \ref{fig:no_rad_fiducial_radial} and
\ref{fig:no_rad_mass_flux_tcool}, and the discussions in
Appendix \ref{sec:scaling-derivation}, we conclude that, in
the regions where hard X-ray emission is dominated by
bremsstrahlung, density may be approximated by the scaling
$\rho\propto r^{-7/4}$, and temperature as
$T\propto r^{-1}$.  Due to the lack of treatment of Compton
cooling in the simulations, we cannot expect those scaling
relations of ``non-conservative'' quantities, such as
temperature and density, to still hold in the IC cooling
dominated region when properly treated.  We thus define the
power-law indices $\alpha$ and $\beta$, for temperature and
density respectively, and leave both adjustable in the IC
dominated region and the density profile in the
bremsstrahlung dominated region $\eta$ also adjustable,
\begin{equation}
  \label{eq:scaling-result}
  T \propto
  \begin{cases}
    r^{-1}\ , & \quad r > r_\tran\ ,\\
    r^{-\alpha}\ ,& \quad r \leq r_\tran\ ;
  \end{cases}
  \quad
  \rho \propto
  \begin{cases}
    r^{-\eta}\ ,& \quad r > r_\tran\ ,\\
    r^{-\beta}\ ,& \quad r \leq r_\tran\ .
  \end{cases}
\end{equation}
Based on the discussion given in Appendix
\ref{sec:scale-ic}, at low $L_\uv$, we use $\alpha = 1$,
while $\alpha + \beta \leq 3/2$ for high $L_\uv$
cases. Also, in Appendix \ref{sec:scale-ff}, we derive that
$\eta\simeq 7/4$ for the situations with low feeding
rate. We note that the profile introduced by equation
\eqref{eq:scaling-result} is applied to extrapolate the
simulation into the central coronal region, where inverse
Compton process possibly contributes significantly to the
overall X-ray emission of {\it individual} AGNs, especially
those with low accretion rates, as we will elaborate in the
rest of this subsection.

The value of bremsstrahlung-inverse Compton transition
radius $r_\tran$ is defined in Equation \eqref{eq:r-tran} in
the Appendix \ref{sec:scaling-derivation}. For
normalization at $r>r_\tran$ outside the IC dominant region,
we use the virial temperature for $T$ as in equation
\eqref{eq:Tvir} and electron density $n_e$ as (both of which
are consistent with simulations)

\begin{equation}
  \label{eq:ne-profile}
  n_e = n_0 \left( \dfrac{r}{3\times 10^3 r_\sch}
  \right)^{-\eta}\ ,\quad r > r_\tran\ . 
\end{equation}
We adjust the normalization of the power-law relations in
the $r\leq r_\tran$ region, so that the $T$ and $n_e$
functions are continuous at $r=r_\tran$.  Also, a ``cap''
temperature is assigned based on Equation
\eqref{eq:T-cap-ic} in the IC dominated region, 
considering cooling and heating balance.

The density and temperature profiles are applied to
calculate the hard X-ray spectra, using a code that we have
developed and thoroughly tested to compute IC scattering and
resultant spectra from first-principle Monte Carlo
simulations (see Appendix \ref{sec:mcic} for details).  This
three-dimensional IC scattering code, called ICode, is to be
made publicly available immediately with access information
provided in Appendix \ref{sec:mcic}.  For the UV source from
disk emission, we adopt single-temperature black body
spectra for simplicity without loss of essential physical
features and observable output, whose peak of distribution
in photon energy is adjustable but is set to be $15\ \eV$
\citep[e.g.][]{1993Haardt} for our calculation presented.
The source is designed as a standard thin disk, whose inner
edge is set to be consistent with the UV luminosity
\citep[see][]{2009tomsick}. At each radius on the disk, the
rate of the number of photons emitted per unit area is in
accordance with the effective temperature profile standard
$\alpha$-disk model \citep[e.g.][]{1973Shakura}.
In some cases where the AGN UV/optical luminosity emanating
from the disk is high, $r_\tran$ may encroach into the
bremsstrahlung regions in current simulations.  For those
cases, we also calculate the bremsstrahlung spectra
analytically with analytically obtained density and
temperature profiles.
We will also explore the impact of the value of $n_0$
defined in Equation \eqref{eq:ne-profile}.

\begin{figure}[h!]
\centering
\vspace{0in}
\includegraphics[width=5in, keepaspectratio]{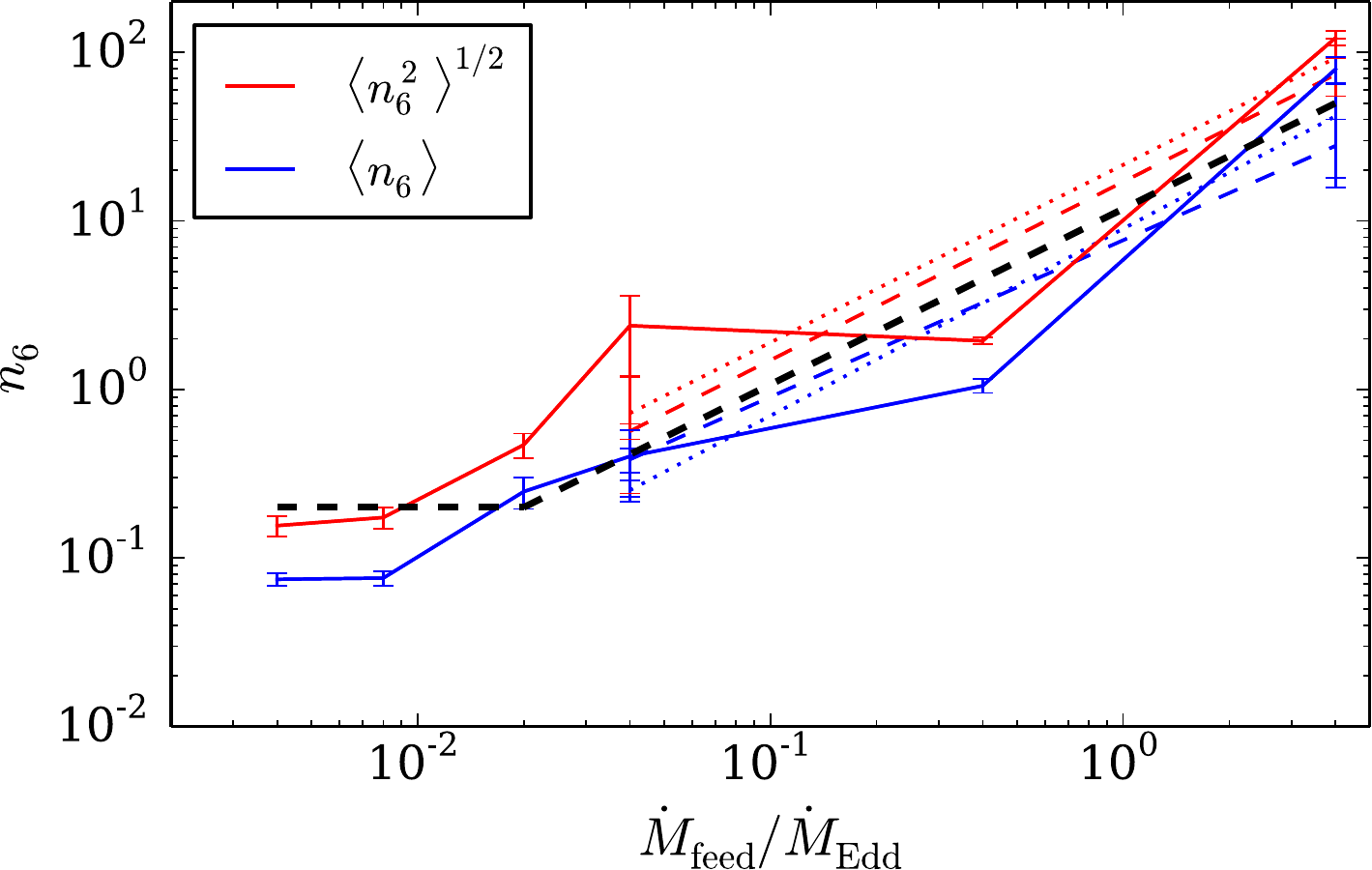}
\vspace{0.0cm}
\caption{ Shows the dependence of $n_6$ on feeding gas
  feeding rate and angular momentum $\jexp/j_0$.  The solid
  lines have $\jexp/j_0 = 10^{-3}$, dotted lines have
  $\jexp/j_0 = 10^{-2}$, and dashed lines have
  $\jexp/j_0 = 10^{-1}$.  The blue lines show $n_6$ that is
  volume averaged, while the red line is the
  root-mean-squared value of
  $n_6$. %which is more relevant to bremsstrahlung emission.
  Errorbars represent the interquartiles.  The black heavy
  dashed line indicates the overall trend of $n_6$'s feeding
  rate dependency.  }
\label{fig:n6_j_mfeed}
\end{figure}

The density of shock heated gas in the central region,
$n_6\equiv n_0/(10^6\ \cm^{-3})$ at
$3\times 10^3r_{\rm sch}$, depends on the feeding rate and
angular momentum of gas at the outer boundary.  Here we will
express gas feeding rate $\mfeed$ in terms of $\mdot_\edd$,
the ``Eddington'' feeding rate for the SMBH, defined as,
\begin{equation}
  \label{eq:mfeed-edd}
   \mdot_\edd = \dfrac{L_\edd}{\alpha_{\rm rad} c^2}\ ,
\end{equation}
where $\alpha_{\rm rad}\equiv L/(\mdot c^2)$ is the
radiative efficiency.  Typically,
$\alpha_{\rm rad} \simeq 0.1$, which gives
$\mdot_\edd \simeq 2.2 M_\odot/\yr$ for
$M_\bh = 10^8 M_\odot$.
%All those low $\mfeed$ runs have $N_\feed = 1$.  
In Figure \ref{fig:n6_j_mfeed}, we show how $n_6$ varies
with different $\jexp$ and $\mdot_\feed$.  It is seen that,
at a given, relatively low $\mdot_\feed$, $n_6$ does not
sensitively depend on $\jexp$, as seen earlier in Figure
\ref{fig:no_rad_emis_spec_compare}.  On the other hand,
$n_6$ increases with increasing $\mfeed$, albeit with a rate
that is sublinear.  At higher $\mfeed/\mdot_\edd$ beyond the
range shown, we find that $n_6$ flattens out and we
attribute the high-end ``saturation'' of $n_6$ to
quadratically-increasing cooling rate with density, which
causes progressively more ``dropout'' of coronal materials.
At the low end of $\mfeed/\mdot_\edd$
($\lesssim 4\times 10^{-3}$), it is seen that $n_6$ also
flatten out.  This is because cooling time starts to
significantly exceed the dynamical time (see also Equation
\ref{eq:tff} and \ref{eq:tcomp}) such that a hot corona may
be ``maintained" nearly independent of $\mfeed$ at low
$\mfeed$ end.  With the results in Figure
\ref{fig:n6_j_mfeed}, if we express
$L_\uv/L_\edd \simeq (\alpha_{rad}/0.1) \mfeed/\mdot_\edd$,
one then obtains a relatively constrained range of $n_6$ for
a given $L_\uv/L_\edd$.  Typically, in the regime that
$\mfeed/\mdot_\edd\gtrsim 10^{-1}$, $n_6$ is seen to be
$\sim 10^0$; below $\mfeed/\mdot_\edd\gtrsim 10^{-1}$, $n_6$
slowly approaches $\sim 10^{-1}$.

\begin{figure}[h!]
\centering
\vspace{0.0cm}
\includegraphics[width=6.0in, keepaspectratio]
{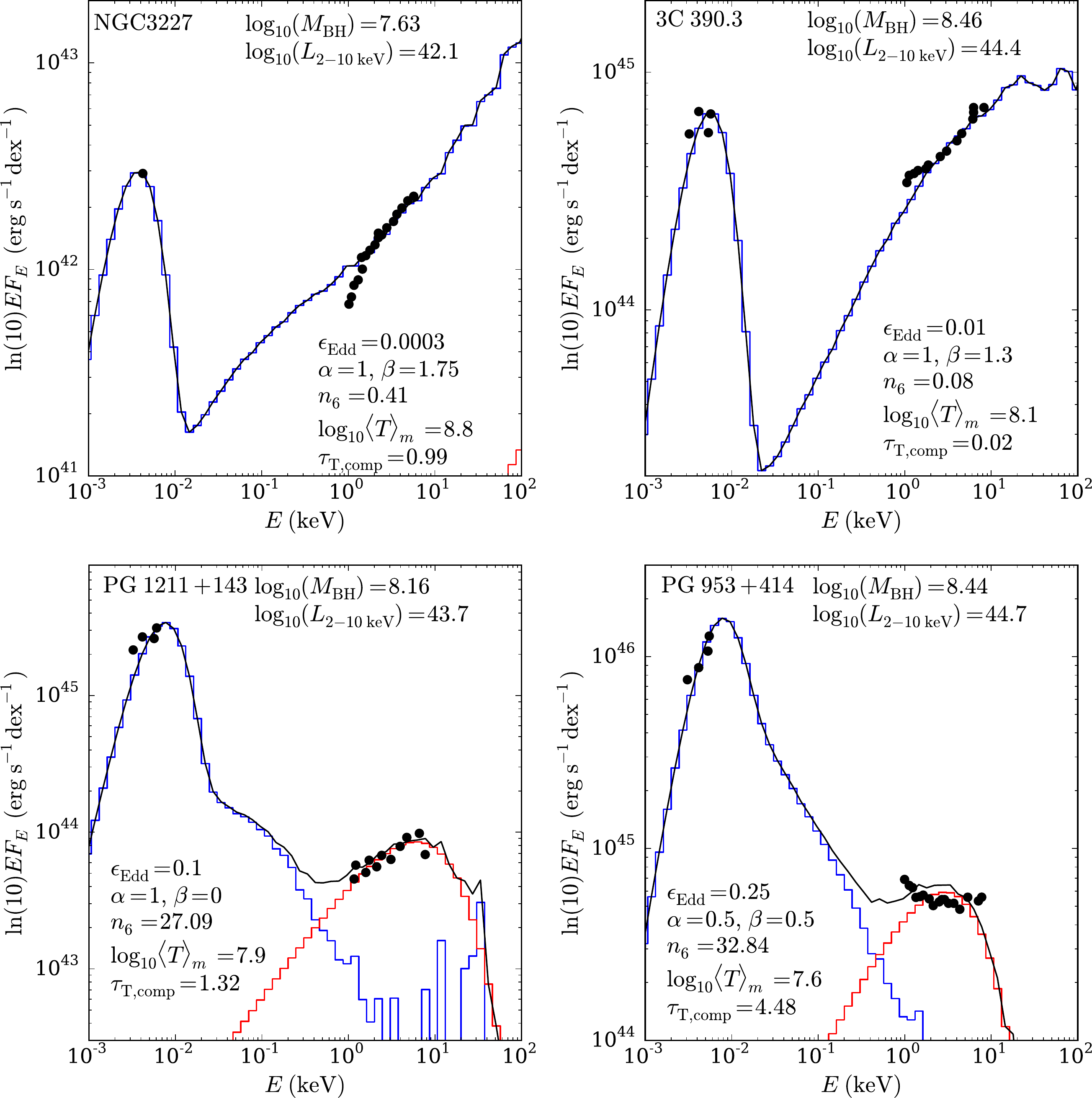}
\vspace{0.0cm}
\caption{Composite UV and X-ray spectra, consisting of
  original UV, inverse Compton and bremsstrahlung
  components. In each panel, the original UV and inverse
  Compton X-ray component generated by Monte Carlo
  simulation is shown in blue stepped curve; analytically
  computed bremsstrahlung contribution is indicated by red
  stepped curve.  The total spectral intensity, which is the
  sum of blue and red curves, is shown by the black curve.
  In each case, the information for the observed object
  (object name, SMBH mass and X-ray luminosity in the
  $\bandkeV{2}{10}$ band) is indicated at the top of the
  panel, while the input model parameters (Eddington ratio
  for the disk luminosity, temperature profile slope
  $\alpha$, density profile slope $\beta$, normalization
  density $n_0$) are indicated in lower part of the
  panel. The derived parameters (total radially integrated
  IC optical depth $\tau_{\rm IC}$ and optical
  depth-weighted temperature of the IC region
  $\langle T_\comp \rangle_\tau$) are also indicated for
  those cases whose X-ray emission is dominated by IC. The
  observed data points (without errorbars) are shown as
  dots.  }
\label{fig:spec_fit_individual} 
\end{figure}

To show the versatility and capability of our model, we use
four observed systems spanning a wide range in spectral
shape \citep[data are from][]{2009vasudevan}.  In our model,
the temperature of black body source of seed photons from the
disk is set by observations according to
\citet[][]{2009vasudevan}.  In this case, we simply vary
$\alpha$ and $\beta$ manually until we arrive at a
reasonable fit.  The results obtained are based on detailed
IC scattering processes using ICode.  It is noted that our
calculations of IC processes are performed in three
dimensions, albeit with the spherically symmetric geometry
for the IC scattering region and the disk geometry for the
UV emitting regions in this case.  The fitting process also
ensures that the Eddington ratio and the luminosity in
$\bandkeV{2}{10}$ both match those of the observed system in
question.  Since our model is constructed based on
simulation results and physical constraints, the arrived set
of model parameters is physically attainable.

Detailed comparisons are shown in Figure
\ref{fig:spec_fit_individual}.  The top two panels show two
cases with hard spectral shape, where the hard X-ray
luminosity is comparable to, or exceeds, that of the UV
bump, whereas the bottom two panels show two cases with
``normal'' X-ray to UV luminosity ratio in the range
$1-10\%$.  Three points are worth noting.  First of all, our
physical model can easily accommodate the observed variety
of spectral shapes.  Second, comparing the top panels with
the bottom panels, we see a trend: the relative IC
contribution to hard X-ray emission increases, as the UV
luminosity from the disk decreases.  The physical reason for
this, at zero-th order, is that a low UV radiation bath
allows for the existence of an enlarged IC region (i.e., a
larger IC optical depth) and a higher mean temperature in
the IC region.  The IC optical depth determines the
probability distribution function of the number of
scatterings (including multiple scattering events), while
the temperature determines the energetics (i.e., energetic
gain of photons) of scattering events.  Third, in a
physically attractive sense, our ``self-consistent'', coupled
treatment of IC processes and central UV source
distinguishes itself from the usual model with the
assignment of the putative hot corona that is separate from
the property of the central UV source.  This removes the
arbitrariness of the hand-picked coronal gas and places
physically attainable models within a limited parameter
space.

\begin{figure}[h!]
\centering
\vspace{0in}
\includegraphics[width=6in, keepaspectratio]
{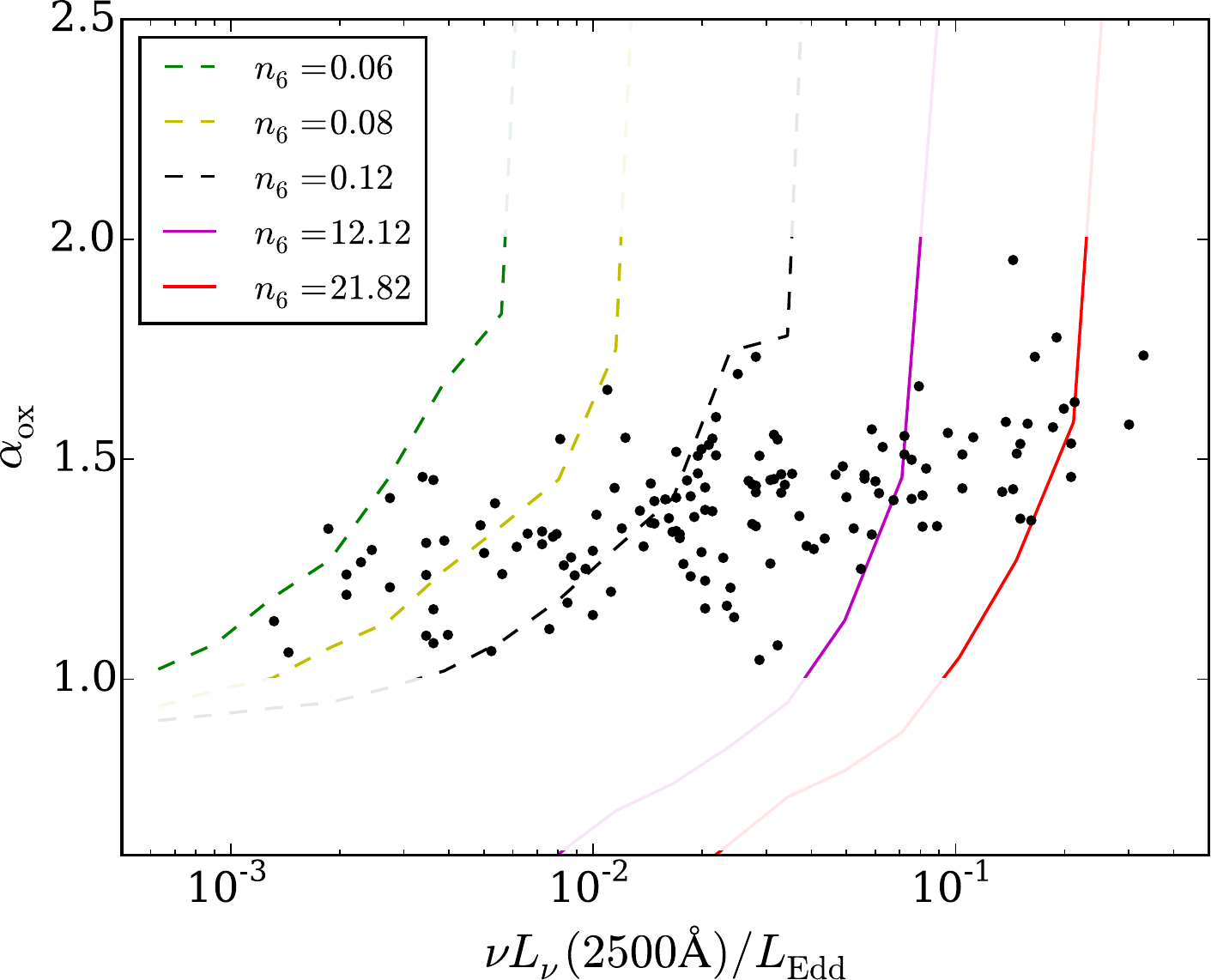}
\vspace{0.0cm}
\caption{Shows our theoretical optical to X-ray spectral
  index $\aox$ as a function of optical/UV luminosity
  $[\nu L_\nu(2500\ \AA)/L_\edd]$ for five different values
  of $n_6\equiv n_0/(10^6\ \cm^{-3})$.  Two sequences are
  shown: dashed lines have $\alpha = 1$ and $\beta = 1.2$,
  and solid lines $\alpha = 1$ and $\beta = 0.5$.  To
  compare reasonably, we convert the luminosity of both
  theoretical and observational data into
  $\nu L_\nu(2500\ \AA)/L_\edd$, which is a dimensionless
  that already takes the variation of observed $M_\bh$ into
  account.  Observational data points, shown as black
  points, are from \citet{2010Lusso}.  The regions that are
  not plausible in terms of parameter space constraint due
  to the correlation between $n_6$ and UV luminosity (i.e.,
  a narrow range of $n_6$ at a given UV luminosity) shown in
  Figure \ref{fig:alpha_corr} are shown as lower and upper
  ``faint'' segments of each line.}
\label{fig:alpha_corr}
\end{figure}

Let us now go beyond just anecdotal evidence, by comparing
with observations statistically and systematically.  
To do that, we use $\aox$, the optical to X-ray spectral index,
with the ``standard'' definition \citep[e.g.][]{2010Lusso}
of
\begin{equation}
  \label{eq:alpha_ox}
  \aox \equiv - \dfrac{ \log_{10}(L_{2\ \keV}/L_{2500\ \AA})
  } {2.605}\ .
\end{equation}
\noindent
We construct our theoretical models with two different sets
of power-law indices: $(\alpha,\ \beta) = (1,\ 1.2)$ (dashed
lines in Figure \ref{fig:alpha_corr}), and
$(\alpha,\ \beta)=(1,\ 0.5)$ (solid lines in Figure
\ref{fig:alpha_corr}).  The chosen values of
$n_0\equiv n_6 \times 10^6\ \cm^{-3}$ span the range found
in Figure \ref{fig:n6_j_mfeed}.  For a given $n_6$, we only
vary the Eddington ratio $\epsilon_\edd$ of the UV photon
source.

Figure \ref{fig:alpha_corr} shows our theoretical optical to
X-ray spectral index $\aox$ as a function of optical/UV
luminosity $[\nu L_\nu(2500\ \AA)/L_\edd]$ for five
different values of $n_6\equiv n_0/(10^6\ \cm^{-3})$.
Comparing the four theoretical curves in Figure
\ref{fig:alpha_corr}, as expected, as $n_6$ increases, with
all other parameters fixed, $\aox$ decreases (i.e., spectrum
becomes harder) at a given Eddington ratio (for UV/optical
luminosity).  This is because a higher $n_6$ gives rise to a
higher IC optical depth hence a larger hard X-ray
luminosity.  Along each curve, as the Eddington ratio (for
UV/optical luminosity) increases, $\aox$ increases (i.e.,
spectrum becomes softer).  This is because a higher
Eddington ratio reduces the IC region hence a lower IC
optical depth and more importantly a lower temperature of
the IC region, in combination resulting in a lower X-ray
luminosity.  Intriguingly and perhaps profoundly, this
simple model with a reasonable range of $n_0$ that is
consistent with Figure \ref{fig:alpha_corr}) can account for
the observed data points exceedingly well.

More specifically, the trend of our model curves going from
lower left to upper right result in, naturally, two
``deserts'' on the upper-left and lower-right corners, seen
in the observational data.  If the model prediction for the
desert in the upper-left corner holds up, the implication is
that a lower value of $n_6\sim 0.1$ is required, as our
simulations suggest, with small but non-negligible AGN
activities of $\nu L_\nu(2500\ \AA)/L_\edd\ge 10^{-3}$ fed
by a commensurate gas inflow.  Similarly, if the model
prediction for the desert in the lower-right corner is
correct, it would imply an upper limit on $n_6\sim 30$,
which may either be accounted for due to cooling saturation
or limitations on gas inflow gas around SMBHs, also
suggested by our simulations.  These two limits are
consistent with the examples seen in Figure
\ref{fig:n6_j_mfeed}.  Moreover, the range of observed
$\aox\sim 1-2$ is naturally explained due to the correlation
between $n_6$ and UV luminosity, a narrow range of $n_6$ at
a given UV luminosity, shown in Figure \ref{fig:alpha_corr}.
We note that there are some degeneracies between $n_0$ and
the choices of temperature and density profiles.
Nonetheless, our model is being set apart from other models
in its physical simplicity, a multitude of predictive power
and falsifiability.

\section{Discussion and Conclusions}
\label{sec:summary}

We perform three-dimensional hydrodynamic simulations,
covering the spatial domain $\sim 10^2 r_\sch$ to $2~$pc
around the central supermassive black hole of mass
$10^8\msun$, with detailed radiative cooling processes and gravity.
Analysis shows that, for a realistic range of gas feeding
rates from large scales ($\sim 2~$pc), gravitational shock
heating is the dominant heating process, resulting in a
significant amount of high temperature gas in the inner
$\le 10^4 r_\sch$ coronal region above the accretion disk
with its radiative cooling time scale exceeding
gravitational heating time scale.  We show that the
composite bremsstrahlung emission spectrum due to coronal
gas of various temperatures from our generic simulations are
in reasonable agreement with the overall ensemble spectrum
of AGNs and hard X-ray background in the $\bandkeV{5}{300}$
range. %(Figure \ref{fig:no_rad_spec_observe}).
This indicates that most of the hard X-ray radiation from
AGNs can be accounted for by the bremsstrahlung emission of
the gravitationally shock heated coronal gas.

We then combine the simulation results with a
post-simulation analysis that includes a treatment of the
inverse Compton processes (up-scattering of soft UV photons
produced by the accretion disk), using a newly developed
code from first-principle Monte Carlo simulations, (see
Appendix \ref{sec:mcic} for details of our newly developed
code that is made publicly available).  One of the most
attractive features of our model is that the hot coronal gas
in the central inverse Compton region is a significantly
constrained and integrated part of the gas feeding process,
rather than an ad hoc separate component. Further, the
property (density and temperature radial profiles) of the
gas in the inverse Compton region is dependent on the
luminosity of the accretion disk.  We show that the combined
modeling can readily account for the wide variety of AGN
spectral shape, which can be understood physically in simple
ways.  One natural outcome, one of the most salient features
of our model, is an anti-correlation between SMBH accretion
disk luminosity and spectral hardness: as the luminosity of
SMBH accretion disk decreases, the hard X-ray luminosity
increases relative to the UV/optical luminosity.  This is
because, as the luminosity of SMBH accretion disk decreases,
the radius of the region whose electrons are not cooled by
UV photons decreases, resulting in an increase in the energy
of scattering electrons.  We show that this general trend
not only is borne out in individual observed AGNs but also
explains the spectral hardness--Eddington ratio relation
observed.  Our model suggests two ``deserts'' of AGNs, with
either low-luminosity and soft spectral index, or
high-luminosity and hard spectral index, that may be
verifiable.  Moreover, the range of observed $\aox\sim 1-2$
is naturally explained due to a relative tight correlation
between the coronal gas density in the central region and
AGN UV luminosity.
% (that is, there is a narrow range of $n_6$ at a given UV
% luminosity.)

\vskip 0.5cm

Computing resources were in part provided by the NASA High-
End Computing (HEC) Program through the NASA Advanced
Supercomputing (NAS) Division at Ames Research Center.  The
research is supported in part by NSF grant AST-1108700 and
NASA grant NNX12AF91G.

\appendix

\section{Scaling Relations and Constraints of Physical
  Quantities }
\label{sec:scaling-derivation}

We start our derivations by assuming a steady state.
Consider a thin spherical shell at radius $r$ with thickness
$\delta r$, and volume $4\pi r^2 \delta r$.  At the outer
and inner boundaries of this shell, mass and energy flow in
and out, respectively.  In general, within the shell, there
are influx and outflux of mass, along with some material
cooling and ``dropping out'' from the hot phase (to join the
disk).  Energetically, thermal energy is injected by
gravitational acceleration and subsequent thermalization,
along with advected thermal energy and cooling.  In a steady
state, the net change of both mass and energy within the
shell is zero.

\subsection{Energy Balance and Scaling Relations in
  Bremsstrahlung Dominated Region}
\label{sec:scale-ff}

In this subsection, we derive the scaling relation of
density and temperature in the region where the dominating
cooling mechanism is bremsstrahlung.

We denote the hot gas ``drop-out'' rate as
$(\d \mdot / \d r)_\mathrm{drop}$.  Conservation of mass in
this shell says,
\begin{equation}
  \begin{split}
    0 = \dfrac{\Delta M}{\d t} = \mdot (r + \delta r) -
    \mdot (r) - \left( \dfrac{\d \mdot}{\d r}
    \right)_\mathrm{drop} \delta r = \left[ \dfrac{\d
        \mdot}{\d r} - \left( \dfrac{\d \mdot_{}}{\d r}
      \right)_\mathrm{drop} \right] \delta r\ ,
  \end{split}
\end{equation}
which then gives the net mass flux through the boundaries
equal to the drop-out rate,
\begin{equation}
  \label{eq:mass-conserve}
  \dfrac{\d \mdot}{\d r} = \left( \dfrac{\d
  \mdot}{\d r} \right)_\mathrm{drop}\ .
\end{equation}
Next, we consider the energy budget within the radial shell,
where its temporal change has three different components:
\begin{equation}
  \label{eq:eng-balance}
  \Delta E = \Delta E_\mathrm{flow} + \Delta
  E_\mathrm{grav} + \Delta
  E_\mathrm{rad} \ , 
\end{equation}
where $\Delta E_\mathrm{flow}$ is the amount of energy that
travels with mass flow, $\Delta E_\mathrm{grav}$ represents
gravitational heating, and $\Delta E_\mathrm{rad}$ accounts
for the radiative cooling.  The $\Delta E_\mathrm{flow}$ term
can be expressed in terms of temperature gradient using
Equation \eqref{eq:mass-conserve} as,
% and the equipartition theorem ($\mu$ is the relative
% molecular mass with electrons taken into account),
\begin{equation}
  \begin{split}
  \label{eq:de_flow}
  \dfrac{\Delta E_\mathrm{flow}}{\d t}
  % & =\dot{U} (r + \delta r) - \dot{U} (r) - \left(
  %   \dfrac{\d
  %   \dot{U}}{\d r} \right)_\mathrm{drop} \delta r
  % //
  = \dfrac{3}{2} T(r + \delta r) \dfrac{\mdot (r + \delta
    r)}{\mu m_p} - \dfrac{3}{2} \kb T (r) \dfrac{\mdot
    (r)}{\mu m_p} - \dfrac{3\kb T(r)}{2\mu m_p} \left(
    \dfrac{\d \mdot}{\d r} \right)_\mathrm{drop} \delta r
  % \\
  & = \dfrac{3 \kb}{\mu m_p} \mdot (r) \dfrac{\d T}{\d r}
  \delta r\ .
\end{split}
\end{equation}
For the gravitational heating component, we express it in
terms of mass flux,
\begin{equation}
  \begin{split}
  \label{eq:de_grav}    
  \Delta E_\mathrm{grav} & = \left[ \left( \dfrac{G
        M_\mathrm{BH}}{r^2} \right) 4 \pi r^2 \delta r \rho
    (r) \right] (| \langle v_r \rangle | \d t) = \left(
    \dfrac{G M_\mathrm{BH}}{r^2} \right) \dot{M \ } \delta r
  \d t \ .
\end{split}
\end{equation}
where we use
$\mdot(r) = 4\pi r^2 \rho |\langle v_r \rangle| $, and
$\langle v_r \rangle$ is the mass-weighted radial velocity
of the shell.

\begin{figure}[h!]
\centering
\vspace{0.0cm}
\includegraphics[width=6.3in, keepaspectratio]{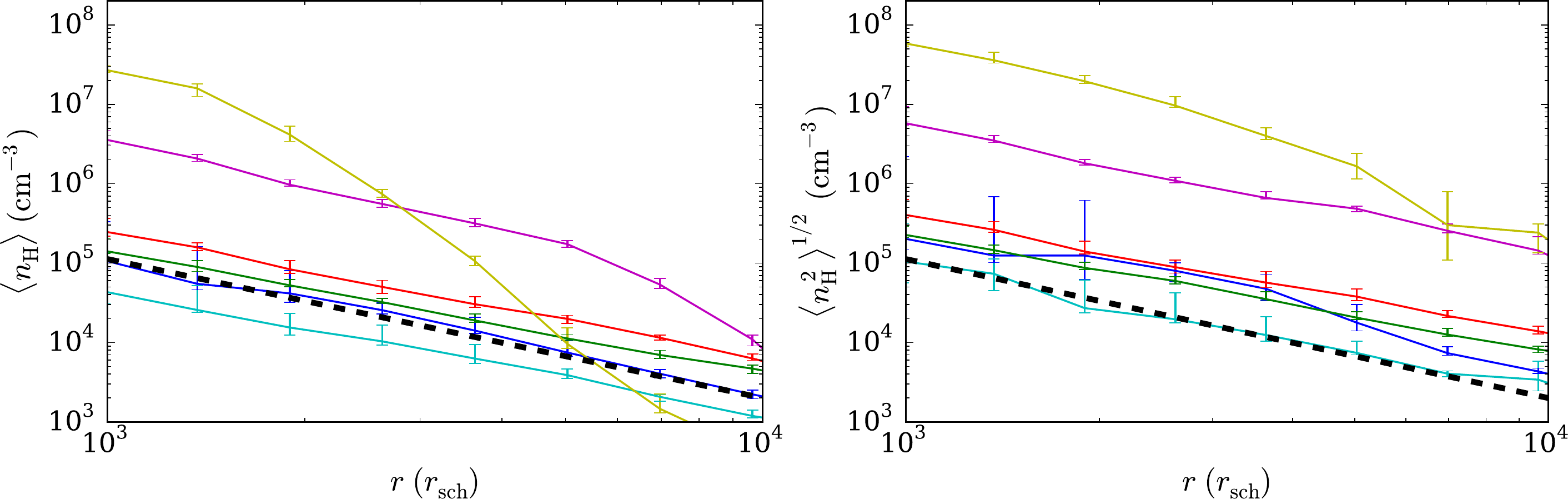}
\vspace{0.0cm}
\caption{ Shows density (left panel:
  $\langle n_\mathrm{H}\rangle$; right panel:
  $\langle n_\mathrm{H}^2\rangle^{1/2}$) profile for runs
  with different feeding rates $\mdot_\feed/\mfeedn$ (cyan:
  0.001, Run 8; blue: 0.002, Run 7; green: 0.005, Run 6;
  red: 0.01, Run 5; magenta: 0.1, Run 4; yellow: 1, Run 3).
  % Left panel shows the density profile of only the infalling
  % gas.  Right panel shows the density profile of all gas.
  The heavy dashed line shows the power index $\beta=7/4$.
}
\label{fig:n_feed} 
\end{figure}

Based on the bremsstrahlung emissivity per unit frequency in
Equation \eqref{eq:jff-emission-spectra} and using the Gaunt
factor equal to unity, the total bremsstrahlung emissivity
is obtained by an integration with respect to $\nu$ is,
\citep[e.g.,][]{2011Draine}
\begin{equation}
  \epsilon_\ff = \int_0^\infty \d \nu j_\ff \simeq C_\ff
  \rho^2 T^{1/2}\ .
\end{equation}
Here $C_\ff$ is a constant.
% that is related to Equation
% \eqref{eq:jff-emission-spectra}.
We note that the Gaunt factor correction is negligible for
the integrated emissivity $\epsilon_\ff$ in hard X-ray
emitting plasma (e.g. the correction factor is only $1.09$
for a $10^9\ \K = 86\ \keV/\kb$ gas).  Hence,
the component of bremsstrahlung energy loss is,
\begin{equation}
  \label{eq:de_rad_ff}
  \left( \dfrac{\Delta E_\mathrm{rad}}{\d t} \right)_\ff = -
  4\pi r^2 \epsilon_\ff \delta r = - 4\pi r^2 C_\ff \rho^2
  T^{1/2} \delta r\ . 
\end{equation}
Combining Equations \eqref{eq:de_flow}, \eqref{eq:de_grav}
and \eqref{eq:de_rad_ff} with Equation
\eqref{eq:eng-balance}, we obtain,
\begin{equation}
  \label{eq:ff-scaling-tot}
  0 = \dfrac{\Delta E}{\d t \delta r} = \dfrac{3 \kb}{2\mu
    m_p} \mdot \dfrac{\d T}{\d r} + \dfrac{G
    M_\mathrm{BH}}{r^2} \mdot  - 4 \pi r^2 C_\ff \rho^2
  T^{1/2} \  . 
\end{equation}
Using this relation and Equation \eqref{eq:Tvir},
\eqref{eq:ff-scaling-tot} becomes
\begin{equation}
  \label{eq:ff-scaling-rho}
  r^2 \rho^2 T^{1 / 2}  \propto \mdot r^{- 2} \ ;\quad \rho
  \propto (r^{- 3} T^{- 1 / 2})^{1 / 2} \propto  r^{- 7 / 4
    + \gamma / 2} \  .  
\end{equation}
We utilize the results found in our simulations to constrain
$\gamma$ in $\mdot(r) \propto r^\gamma$.  We find that
$\gamma \simeq 0$ at the low feeding rate limit, which
yields $\beta=7/4$.  At at high feeding rate limit,
$\gamma \simeq 1/2$, yielding $\beta = 3/2$. %  We make a
% subtle but important distinction about the gas at a given
% radius in order to properly assess the density profile of
% infalling gas for low feeding rate cases.  We separate
% ``infalling" gas from ``sitting" gas, where ``infalling" gas
% is separated out as the component that has larger radial
% velocity than the tangential component.  In the left panel
% of of Figure \ref{fig:n_feed} we see that the density
% profile of the ``infalling" gas follows approximately the
% $\beta=7/4$ power-law slope in the inner region, whereas the
% ``sitting" gas behaves differently.  Note that the amount of
% ``sitting" gas increases relatively with decreasing density,
% as expected, since cooling time increases with decreasing
% density.
% The static part on the right panel, on the other hand,
% shows steeper increase as $r$ becomes smaller, indicating
% that there are more gas than the infalling part staying in
% the coronae.
We plot the density profile for the hot ($T > 10^6\ \K$) gas
in the inner region
% (viz. only the part of gas that is emitting in X-ray
% comes into statistics here, which makes the plots
% different from Figure \ref{fig:no_rad_mass_flux_tcool}),
both arithmetic average and root-mean-squared average
through $4\pi$ solid angle, in Figure
\ref{fig:n_feed}. Compared with the guiding line showing
$\beta = 7/4$ slope, the density profile in the central
regions agrees nicely with our derivations.

\subsection{Transition from Bremsstrahlung to Inverse
  Compton}
\label{sec:trans-ff-ic}

\begin{figure}[h!]
\centering
\vspace{0.0cm}
\includegraphics[width=5.3in, keepaspectratio]
{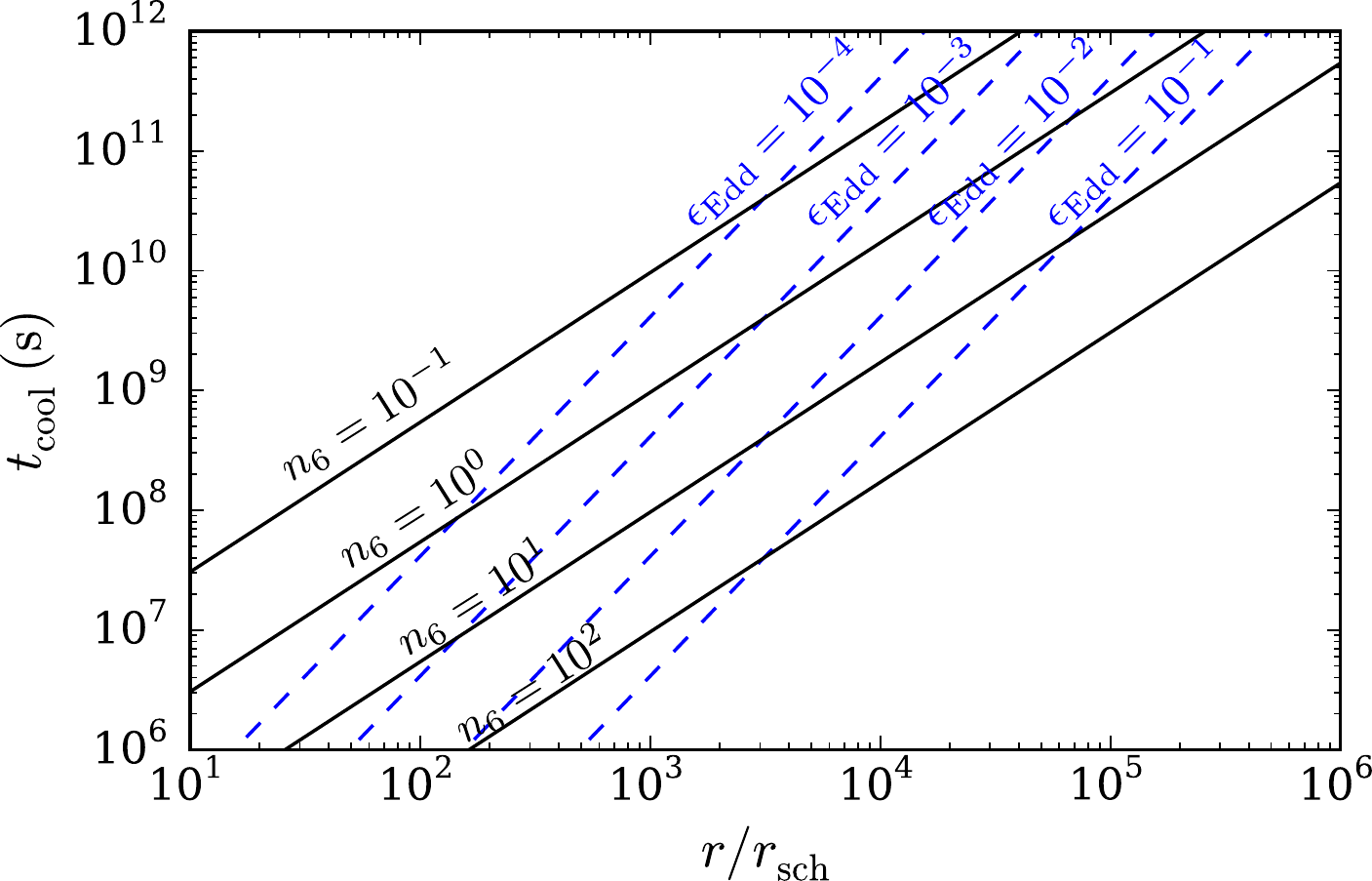}
\vspace{0.0cm}
\caption{Compares inverse Compton cooling time [blue dashed
  lines; see equation \eqref{eq:tcomp}] and free-free
  cooling time [black solid lines, see equation
  \eqref{eq:tff}] as functions of radius. $\eta = 7/4$ is
  taken here for radial density profile, and other necessary
  parameters are labeled in the figure near each line. }
\label{fig:compare_t_cool} 
\end{figure}

We now derive the expression of another important quantity,
the optical depth of the central inverse Compton region.  We
define the ``transition radius'', $r_\tran$, such that
inverse Compton cooling dominates at $r<r_\tran$ and
bremsstrahlung cooling dominates at $r>r_\tran$, and they
are equal at $r_\tran$.  It is easily seen that, using
$\eta = 7/4$, in terms of $r$, the ratio between $t_\comp$
and $t_\ff$ varies as,
\begin{equation}
  \label{eq:tic-tff-ratio}
  \dfrac{t_\comp}{t_\ff} \propto \dfrac{r^2}{r^{5/4}}
  \propto r^{3/4}\ . 
\end{equation}
Setting $t_\comp/t_\ff$ to unity, we obtain,
\begin{equation}
  \label{eq:r-tran}
  r_\tran \simeq 3 \times 10^7r_\sch\ \left(\dfrac{n_0}
    {10^6\ \cm^{-3}}\right)^{-4/3} \left( \dfrac{M_\bh}{10^8
      \msun}\right)^{4/3} \epsilon_\edd^{4/3}\ . 
\end{equation}
In Figure \ref{fig:compare_t_cool}, we illustrate the
comparison between inverse Compton and analytically
extrapolated free-free cooling time scales, under different
luminosity ($\epsilon_\edd$) and feeding ($n_6$)
conditions. With each combination of parameters, the value
of $r_\tran$ is illustrated by a crossing point of an
inverse Compton (blue dashed) line and a free-free (black
solid) line.

This directly leads to the Thomson optical depth of inverse
Compton region, assuming a $n\propto r^{-\beta}$ density
profile,
\begin{equation}
  \label{eq:ic-tau}
  \begin{split}
    \tau_{\T,\comp} & \simeq \sigma_\T n_0 \left(
      \dfrac{r_\tran}{3\times 10^3r_\sch} \right)^{-7/4}
    r_\tran \int_{r_\sch}^{r_\tran} \left(
      \dfrac{r}{r_\tran}
    \right)^{-\beta} \dfrac{\d r}{r_\tran} \\
    & = 0.0033\ \epsilon_\edd^{-1}\left(
      \dfrac{M_\bh}{10^8\msun} \right)^{-1} \left(
      \dfrac{n_0}{10^6\ \cm^{-3}} \right) \times
  \begin{cases}
    \ln\left(\dfrac{r_\tran}{r_\sch}\right)\ ,\quad & \beta
    = 1\ ;\\
    \dfrac{1}{1-\beta}\left[ 1 -
      \left(\dfrac{r_\sch}{r_\tran}\right)^{1-\beta}
    \right]\ ,\quad & \beta \neq 1\ .
  \end{cases}
\end{split}
\end{equation}
Note the anti-correlation between 
$\tau_{\T,\comp}$ and $\epsilon_\edd$.

\subsection{Constraints and Energy Balance in Inverse
  Compton Region}
\label{sec:scale-ic}

Here we derive the scaling relations in the innermost
coronal region, where cooling process is dominated by
inverse Compton. Under the assumption of a steady state,
there are two different types, depending on the
feeding/inflow rate and gas density.

In the first type, gas feeding/inflow rate is low and gas
density in the central region correspondingly low.  In this
case, the gas cooling time is longer than dynamical time,
and hence the gas ``turnover" (i.e., cooling off) rate is
low.  Thus, the gas may be considered to be in
quasi-hydrostatic equilibrium.  The coronal gas in
quasi-hydrostatic equilibrium in the spherical symmetry case
reads,
\begin{equation}
  \label{eq:ic-hydrostat}
  - \dfrac{G M_\bh}{r^2} + \dfrac{1}{\rho} \dfrac{\d p}{\d
    r}  = 0 \ ;\quad
  \dfrac{1}{\rho} \dfrac{\d (\rho T)}{\d r} \propto r^{-2}\ .
\end{equation}
Assuming power-law density and temperature profiles
specified in equation \eqref{eq:scaling-result}, we insert
$\alpha$ and $\beta$ into equation \eqref{eq:ic-hydrostat},
to result in
\begin{equation}
  \label{eq:ic-t-scaling}
  r^{\beta} \dfrac{\d}{\d r} r^{-(\alpha+\beta)} \propto
  r^{-2}\ ;\quad \alpha = 1\ .
\end{equation}
The power-law index for the density profile, under the
quasi-hydrostatic equilibrium condition, is not constrained,
due to the domination of SMBH gravity over coronal gas
self-gravity.

In the second type, gas feeding/inflow rate is high and gas
density in the central region correspondingly high.  In this
case, the gas cooling time is shorter than dynamical time,
and hence the gas is in constant flux.  Although the steady
state assumption may still be valid, the gas is far from
being in quasi-hydrostatic equilibrium.  For this regime, we
constrain the power-law profiles as follows.

For a parcel of hot electrons bathed in a radiation field of
energy density $u_\ph$, the cooling time of the electrons
due to IC processes is \citep[e.g.][]{1979rybicki}
\begin{equation}
\label{eq:tcomp}
\begin{split}
  t_\comp & = {3\over 8}{m_e c\over \sigma_T}{1\over u_\ph} \\
  & = 3.7 \times 10^7\ \s \left({M_\bh\over
      10^8\msun}\right) \left({\epsilon_\edd \over
      0.1}\right)^{-1} \left({r\over 3\times
      10^3r_\sch}\right)^2\ ,
\end{split}
\end{equation}
where $m_e$ is the electron mass, $c$ the speed of light,
$\sigma_T$ the Thompson cross section,
$\epsilon_\edd \equiv L_\uv / L_\edd$ the Eddington ratio
for the disk luminosity.  One important feature to note is
that $t_\comp$ decreases with decreasing radius $r$,
implying a tendency to produce hard X-ray spectrum via IC
processes.  Using Equation \eqref{eq:tcomp}, we deduce the
scaling relation of energy loss rate due to inverse Compton
(the dependencies on %$\mu$,
$M_\bh$ and $\epsilon_\edd$ are absorbed into the factor
$C_\comp$, which is independent of radius)
\begin{equation}
  \label{eq:emis-ic}
  \epsilon_\comp = C_\comp \rho T r^{-2}\ .
\end{equation}
\noindent 
Using the same arguments used earlier in the derivation for
the bremsstrahlung region (\S A.1), we obtain the radiative
energy loss term
\begin{equation}
  \label{eq:de_rad_ic}
  \left( \dfrac{\Delta E_\mathrm{rad}}{\d t} \right)_\comp =
  - 4\pi r^2 \epsilon_\comp \delta r = - 4\pi C_\comp \rho T
  \delta r\ , 
\end{equation}
\noindent 
which, when combined with Equations \eqref{eq:eng-balance},
\eqref{eq:de_flow} \eqref{eq:de_grav} gives
\begin{equation}
 \label{eq:ic-scaling-rho}
  \begin{split}
    \rho \propto T^{-1} \left( \dfrac{3 \kb}{2\mu m_p}
      \dfrac{\d T}{\d r} + \dfrac{G M_\mathrm{BH}}{r^2}
    \right) \mdot\ ;\quad \rho T \propto \mdot r^{-2}\ .
  \end{split}
\end{equation}
If we further assume that the relation
$\mdot \propto r^{1/2}$ (for high feeding rate) still holds
in the inverse Compton domain, we have
\begin{equation}
 \label{eq:ic-scaling-rho}
\alpha + \beta = 3/2\ .
\end{equation}
\noindent 
It is reasonable to assume that the simulation results
without IC cooling may be extrapolated into the IC cooling
region in the high gas feeding rate regime, because cooling
is important and gas dynamics is primarily determined by
gravity and hydrodynamics, not primarily by thermodynamics.
In any case, since IC cooling becomes more important at
smaller radii, which in turn may induce, relatively, more
gas drop-out at smaller radii, it is therefore likely that
$\alpha + \beta \leq 3/2$ holds in realistic situations.

We utilize the inverse Compton cooling and
gravitational heating processes to provide further
constraints.  For a fluid element of mass $\delta m$ falling
towards the central black hole at virial velocity, the
gravitational heating rate is
\begin{equation}
  \left( \dfrac{\Delta E}{\d t} \right)_\grav = F_\grav
  \left( \dfrac{\d r}{\d t} \right)_{\mathrm{virial}} 
  \leq \dfrac{G M_\bh \delta m}{r^2} \left( \dfrac{G
      M_\bh}{r} \right)^{1 / 2} 
  = \dfrac{(G M_\bh)^{3 / 2}}{r^{5 / 2}} \delta m \  . 
\end{equation}
\noindent
Plugging in various numbers, we have the energy gain rate
per unit mass due to gravitational heating
\begin{equation}
  \label{eq:grav-heat-rate}
  \left( \dfrac{\Delta E}{\delta m \d t} \right)_\grav
  \leq 1.02 \times 10^{10} \  \erg\ \s^{-1}\ \g^{-1}\ 
  \left( \dfrac{M_\bh}{10^8 \msun} \right)^{- 1}
  \left( \dfrac{r}{10^3 r_\sch} \right)^{- 5 / 2} \  . 
\end{equation}
In the mean time, the amount of energy that is removed from
this fluid element per unit time is,
\begin{equation}
  \left( -\dfrac{\Delta E}{\d t} \right)_\comp \simeq
  \left( \dfrac{3 \kb T \delta m}{2 m_p} \right) \left(
  \dfrac{3 m_e c}{8 \sigma_\T} \dfrac{4 \pi r^2 c}{L_\edd
  \varepsilon_\edd} \right)^{- 1} \  . 
\end{equation}
Inserting the numbers, we get the inverse Compton energy
loss rate per unit mass,
\begin{eqnarray}
  \left( - \dfrac{\Delta E}{\delta m \d t} \right)_\comp
  = 3.08 \times 10^{11} \  \erg\ \s^{-1}\ \g^{-1}
  \epsilon_\edd \left( \dfrac{M_\bh}{10^8 \msun}
  \right)^{- 1} \left( \dfrac{T}{10^9 \  \K} \right) \left(
  \dfrac{r}{10^3 r_\sch} \right)^{- 2} \  . 
\end{eqnarray}
The ratio is
\begin{equation}
  \left( \dfrac{\Delta E}{\delta m \d t}
  \right)_\grav/\left( - \dfrac{\Delta E}{\delta m \d t}
  \right)_\comp \leq 0.033 \  \varepsilon_\edd^{- 1} \left(
    \dfrac{r}{10^3 r_\sch} \right)^{- 1 / 2} \left(
    \dfrac{T}{10^9 \  \K} \right)^{- 1} \  . 
\end{equation}
This result yields an upper limit on temperature in the
inverse Compton region as a function of Eddington ratio
$\epsilon_\edd$ and radius $r$ by setting the ratio to unity:
\begin{equation}
  \label{eq:T-cap-ic}
  T_{\max, \comp} \lesssim 1.1 \times 10^8 \  \K \ 
  \left({\epsilon_\edd\over 0.1}\right)^{-1} \left(
    \dfrac{r}{3 \times 10^3 r_\sch}
  \right)^{- 1 / 2} \  . 
\end{equation}

\section{ICode: A Monte Carlo Code for Inverse
  Compton Scattering Processes}
\label{sec:mcic}

In order to compute the spectra taking into account the
inverse Compton scattering processes, we have developed a
Monte Carlo code in \verb|C++| with \verb|OpenMP|
shared-memory parallelization, called ``ICode".  ICode
follows directly the transport and scattering processes of
photons from first principles.  ICode is constructed with
modular structures and allows for easy modifications for
physical conditions, such as spatial density and temperature
structures of the scattering plasma or the geometry of the
sources of the soft seed photons.  We emphasize that ICode
does not require any simplification of the geometry of the
scattering medium or photon sources and can handle arbitrary
geometry for both.  We properly treat all relevant regimes,
including the ``tran-relativistic'' high-energy electrons as
well as photons with energy comparable to electron's rest
energy.  We now describe the basic physical steps used to
construct ICode.

As a photon is emitted by the soft photon source, its
four-momentum is assigned in such a way that its energy
obeys the spectral distribution of the source (such as a
single- or multi-temperature Planck distribution) and its
direction is random (we assume that the source is isotropic
everywhere, so that the spatial component of photon's
initial four-momentum is uniformly distributed across the
$4\pi$ solid angle).

Before each scattering, a random number $\tau$ that obeys
exponential distribution with unitary parameter [i.e.
$\tau\sim P_{\exp}(\tau;\ 1)=\e^{-\tau}$] is generated. This
$\tau$ is actually the ``optical depth'' that this photon
travels through before it is scattered by a high-energy
electron. Starting from an initial point $\mathbf{x}_0$,
this photon ``walks'' along the chosen direction of
momentum, until the cumulative optical depth
\begin{equation}
  \label{eq:mcic-tau-travel}
  \tau(\mathbf{x}_1,\ \mathbf{x}_0) \equiv
  \int_{\mathbf{x}_0}^{\mathbf{x}_1} |\d \mathbf{x}|
  \sigma_\kn\left( \dfrac{h\nu}{m_e c^2} \right) n_e (
  \mathbf{x} )\ ,
\end{equation}
reaches the desired $\tau$ (that we just generated randomly
following the prescribed distribution) at some
$\mathbf{x}_1$, where the photon is scattered by an
electron.  In Equation \eqref{eq:mcic-tau-travel},
$n_e(\mathbf{x})$ is the electron number density profile of
the hot plasma, and $\sigma_\kn$ is the Klein-Nishina total
cross section for photon-electron scattering
\citep[e.g.][]{1970blumenthal, 2007srednick}
\begin{equation}
  \label{eq:mcic-kn-tot}
  \begin{split}
    \sigma_\kn (x) & = \frac{3}{8} \sigma_\T f_\kn(x) ,\quad
    x \equiv \frac{h \nu}{m_e c^2} \ ;\\
    f_\kn(x) & \equiv \left\{ \frac{2 [x (x + 1) (x + 8) +
        2]}{x^2 (2 x + 1)^2} + \left( \frac{x^2 - 2 x -
          2}{x^3} \right) \ln (2 x + 1) \right\}\ .
  \end{split}
\end{equation}
Since there may be cases where electron thermal energy
is comparable to $m_ec^2$ (hence up-scattered photons with
energy close to $m_ec^2$ are generated), it is desirable to
include Klein-Nishina correction to account for second or
higher order inverse Compton scatterings. The value of
$h\nu$ is exactly the $0$th component of the photon's
four-momentum.
% We note that the Klein-Nishina correction is important as
% the incident photon's energy approaches $m_e c^2$.

At point $\mathbf{x}_1$, the Lorentz factor $\gamma$ of an
electron that interacts with the photon is another random
variable, which obeys the relativistic Maxwellian
distribution, taking the local electron temperature $T_e$ as
an argument \citep[e.g.][]{1993haardt_b},
\begin{equation}
  \label{eq:mcic-electron-maxwell}
  \gamma \sim P_\e( \gamma,\ \Theta_e) = \frac{1}{\Theta_e
    K_2 (1 / \Theta_e)} \gamma (\gamma^2 - 1)^{1 / 2} \exp
  \left( - \frac{\gamma}{\Theta_e} \right) \ ;\quad 
  \Theta_e \equiv \frac{\kb T_e}{m_e c^2} \ . 
\end{equation}
Here, $K_2$ is the second-order modified Bessel function of
the second kind. The direction of that electron's spatial
momentum is randomly, uniformly distributed over the $4\pi$
solid angle.

The most convenient way to deal with scatterings is to adopt
the appropriate differential cross section in the ``target
rest frame'', i.e., the ``electron rest frame'' (ERF for
short), in which the pre-scattering electron is at rest at
the origin.  Our numerical schemes follows the following
steps, starting from the lab frame (LF for short):
\begin{enumerate}
\item Use a proper Lorentz transformation
  $\Lambda^{\mu}_{\;\nu}$ to obtain four-momentum of the
  electron and photon in ERF;
\item Spatially rotate the system in ERF so that the
  incident photon is going along the $x$-axis;
\item Use the normalized proper differential cross section
  and the Compton scattering formula to obtain the photon's
  post-scattering four-momentum in the rotated ERF;
\item Spatially rotate the system back to obtain the
  photon's momentum in the original ERF;
\item Use the inverse Lorentz transformation to obtain the
  photon's four-momentum after scattering in LF.
\end{enumerate}
We denote the pre-scattering four-momentum of the electron
in the LF, to be decomposed, as (following the standard of
special relativity, the symbol ``$\asymp$'' means ``such a
vector/tensor can be represented by such a matrix''),
\begin{equation}
  p_e^\mu \asymp \gamma m_e c \left( 1, \  \beta_x,
    \  \beta_y, \  \beta_z \right)^\T \ ,
\end{equation}
where $\beta_x=v_x/c$ is the relativistic $\beta$ along
$x$-axis. The Lorentz transformation in Step 1 is written as
a function of
$\boldsymbol{\beta}\equiv (\beta_x,\ \beta_y,\ \beta_z)$,
\begin{equation}
  \label{eq:mcic-lorentz}
   \Lambda^\mu_{\;\nu} (\boldsymbol{\beta}) \asymp
   \left[\begin{array}{cccc}
    \gamma & - \gamma \beta_x & - \gamma \beta_y & - \gamma \beta_z\\
    - \gamma \beta_x & 1 + (\gamma - 1) \frac{\beta_x^2}{\beta^2} & (\gamma -
    1) \frac{\beta_x \beta_y}{\beta^2} & (\gamma - 1) \frac{\beta_x
    \beta_z}{\beta^2}\\
    - \gamma \beta_y & (\gamma - 1) \frac{\beta_x \beta_y}{\beta^2} & 1 +
    (\gamma - 1) \frac{\beta_y^2}{\beta^2} & (\gamma - 1) \frac{\beta_y
    \beta_z}{\beta^2}\\
    - \gamma \beta_z & (\gamma - 1) \frac{\beta_x \beta_z}{\beta^2} & (\gamma
    - 1) \frac{\beta_y \beta_z}{\beta^2} & 1 + (\gamma - 1)
    \frac{\beta_z^2}{\beta^2}
  \end{array}\right]\ .
\end{equation}
It is straightforward to verify that the inverse Lorentz
transformation needed in Step 5 can be obtained by
$\Lambda^\dagger{}^{\mu}_{\;\nu} =
\Lambda^\mu_{\;\nu}(-\boldsymbol{\beta})$.
For the spatial rotation in Step 4, the rotation matrix
$R^\mu_{\;\nu}$ for the four-vector consists of two
sequential Eulerian rotations (the Einstein convention of
summation is used hereafter),
\begin{equation}
  \label{eq:mcic-euler-rot}
  \begin{split}
  R^\mu_{\;\nu}(\Theta,\ \Phi) = T^\mu_{\;\rho}(\Phi)
  S^\rho_{\;\nu}(\Theta)\ ;\quad
  T^\mu_{\;\rho}(\Phi) & \asymp 
  \left[\begin{array}{rrrr}
    1 &  &  & \\
    & \cos \Phi & - \sin \Phi & \\
    & \sin \Phi & \cos \Phi & \\
    &  &  & 1
  \end{array}\right] \ ;\\
  S^\rho_{\;\nu}(\Theta) & \asymp
  \left[\begin{array}{rrrr}
    1 &  &  & \\
    & \sin \Theta &  & - \cos \Theta\\
    &  & 1 & \\
    & \cos \Theta &  & \sin \Theta
        \end{array}\right]\ .
  \end{split}
\end{equation}
Here $\Theta$ and $\Phi$ are the direction cosines of
pre-scattering photon in ERF--they are defined in such a way
that the pre-scattering four-momentum of the photon in ERF
is written as (its pre-scattering energy in ERF is denoted
as $h\nu$),
\begin{equation}
  \label{eq:mcic-4-mom-ph-prescat}
  p'{}_{\ph}^\mu \asymp h \nu \left( 1, \  \sin
    \Theta \cos \Phi, \  \sin \Theta \sin \Phi, \  \cos
    \Theta \right)^\T  \ .
\end{equation}
The inverse rotation matrix needed in Step 2 is obtained by
$R^{-1}{}^{\mu}_{\;\nu}(\Theta,\ \Phi) =
S^{-1}{}^\rho_{\;\nu}(\Theta)T^{-1}{}^\mu_{\;\rho}(\Phi)$,
and the inverses of $S$ and $T$ tensors can be obtained by
transposing their corresponding matrices in Equation
\eqref{eq:mcic-euler-rot}.

\begin{figure}[h!]
\centering
\vspace{0.0cm}
\includegraphics[width=6.3in, keepaspectratio]
{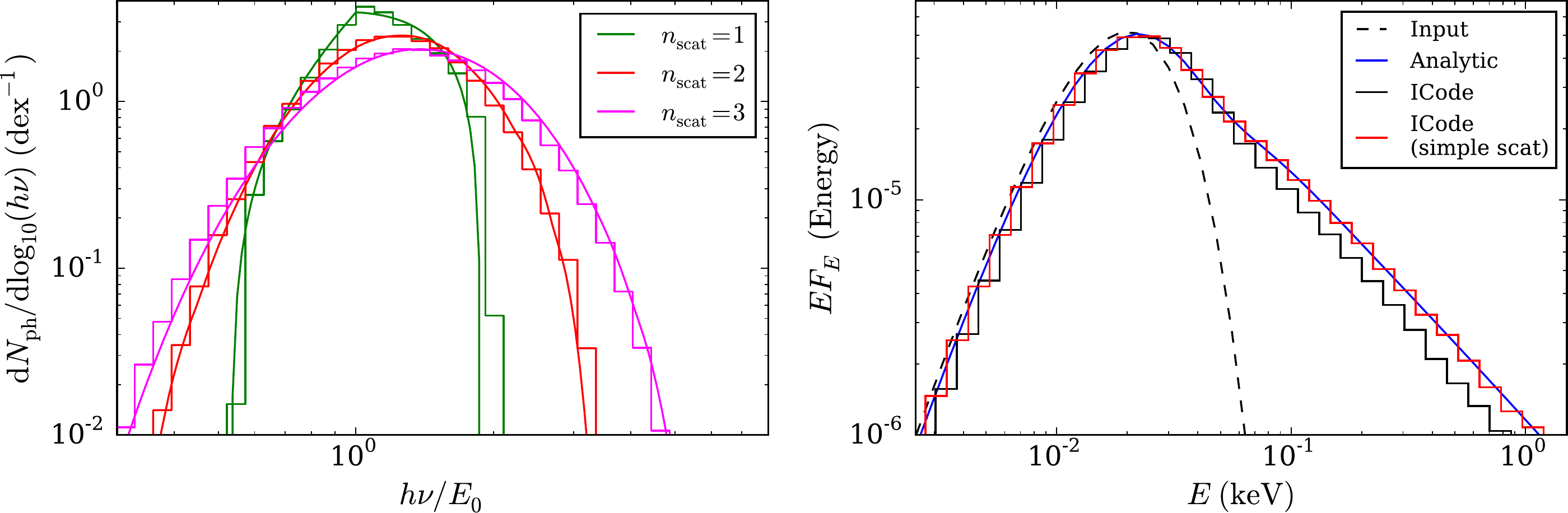}
\vspace{0.0cm}
\caption{Left panel compares ICode simulated inverse Compton
  photon energy distribution shown as step curves with the
  analytic approximation results shown as smooth curves
  based on \citet{1979rybicki}, for one (green curves), two
  (red curves) and three (magenta curves) consecutive
  scatterings.  The initial photon energy is $E_0 = 51\ \eV$
  and electrons have single energy of $25.6\ \keV$
  ($\gamma=1.05$).
  Right panel presents the comparison in a
  more ``realistic'' situation, in which photons have
  initial energy as a black-body spectrum at
  $T = 5.11\ \eV/\kb$ (black dashed curve), injected at the
  center of a spherical cloud at constant electron
  temperature $T_e = 0.1 m_e c^2/\kb = 5.93\times 10^8\ \K$
  and homogeneous electron density
  $n_e = 6\times 10^6\ \cm^{-3}$, with radius $r=0.04\ \pc$.
  Analytic result based on \cite{1979rybicki} and
  \cite{1993haardt_b} is plotted in blue smooth curve. Monte
  Carlo results are presented in step curves, for ordinary
  scattering model (black curve) and ``simple mode''
  scattering model (red curve), respectively.}
\label{fig:mc_single_e_test} 
\end{figure}

For Step 3, the scattering cosine $\mu=\cos\theta$ ($\theta$
is the scattering angle) of the post-scattering photon is
(yet) another random variable that obeys the normalized
Klein-Nishina distribution function,
\begin{equation}
  \mu \sim P_\kn\left( \mu;\ \dfrac{h\nu}{m_e c^2} \right)\ ,
\end{equation}
whose expression reads \citep[see also][] {1970blumenthal,
  2007srednick},
\begin{equation}
  \label{eq:mcic-kn-dist}
  P_\kn \left( \mu; \ \frac{h \nu}{m_e c^2} \right) 
  = \left[ f_\kn \left(\frac{h \nu}{m_e c^2} \right)\right]^{-1}
  P_\comp^2 (\nu, \mu) \left[ P_\comp (\nu, \mu) +
    \frac{1}{P_\comp (\nu, \mu)} - 1 +\mu^2 \right] \ .
\end{equation}
Here, $f_\kn$ is defined in Equation \eqref{eq:mcic-kn-tot},
and $P_\comp$ is the Compton factor, as a function of
incident photon frequency $\nu$ and the scattering cosine
$\mu$,
\begin{equation}
  \label{eq:mcic-compton-factor}
  P( \nu,\mu) \equiv \left[ 1 + \frac{h \nu}{m_e c^2} (1 - \mu)
  \right]^{- 1}\ ,
\end{equation}
which is the post- to pre-scattering photon energy ratio, 
$P_\comp(\nu_0, \mu)=\nu_1/\nu_0$.

With all those steps taken, a photon is scattered to and
travels along a new spatial direction that it just acquired.
Then the above steps are repeated for the scattered photon,
until it leaves eventually the interaction region, at which
point it is collected and binned to obtain the inverse
Compton spectra.
% given initial soft photon spectrum, and the density and
% temperature partial distributions of the hot plasma.

We also introduce a ``simple mode'' to our code, where the
outcome of a single inverse Compton scattering is evaluated
via the approximated formula given in
e.g. \cite{1979rybicki}. This mode ignores such major
complications as electron recoil and anisotropic cross
section, which is a good approximation in the low-energy
regime.

We have thoroughly tested ICode step by step.  We will just
present two non-trivial tests to demonstrate the
verification of the code. The comparison between results
from ICode and analytic calculations is shown in Figure
\ref{fig:mc_single_e_test}.

In the first test problem, shown in the left panel of Figure
\ref{fig:mc_single_e_test}, we inject single-energy photon
at $E_0 = 51\ \eV$ and let the ensemble of photons
experience 1, 2 and 3 inverse Compton scatterings,
respectively, by $\gamma=1.05$ (i.e. kinetic energy
$25.6\ \keV$) single-energy electrons, and compare the
results to analytic calculations based on the approximate
procedures in \citet{1979rybicki}.  This is a strong test on
the ``units" used throughout our code, since the outcome of
Compton scatterings is not convolved with electron or photon
energy distribution functions.  There is a small discrepancy
in the Monte Carlo results tend to be slightly softer for
higher order scatterings compared to the approximated
analytic results, which we attribute to the inaccuracy
introduced in the approximate procedure.  Overall, the
agreement is excellent.

In the second problem, illustrated by the right panel in
Figure \ref{fig:mc_single_e_test}, we put the code to a more
complicated test, where a photon source is embedded at the
center of an isothermal
($T_e = 0.1 m_e c^2/\kb = 5.93\times 10^8\ \K$) and
homogeneous ($n_e = 6\times 10^6\ \cm^{-3}$) spherical
($r=0.04\ \pc$) cloud.  The Thompson optical depth from the
center to the surface of the cloud is $0.49$.  Injected
photons have a black-body spectrum at temperature
$T = 5.11\ \eV/\kb$.  This test case resembles one of the
typical situations that we might encounter when dealing with
AGN corona models.  For comparison, analytic results are
obtained based on the method used in \cite{1993haardt_b}.
The agreement is nearly perfect.  As discussed above, the
small discrepancy between ``simple-mode'' and
``ordinary-mode'' (actually used in ICode) is attributed to
the inaccuracy of simplified treatment of inverse Compton
processes in the ``simple-mode'', which becomes more severe
when photons are scattered multiple times.  In this test,
some photons are scattered multiple times, which is
manifested by the high-end power-law-like tail of the
resulting X-ray photons shown.  In our test, we do not
artificially limit the number of scatterings that a single
photon may encounter.

Given quite thorough tests done, including but not limited
to the two shown above, we conclude that the ICode is ready
to be distributed to the community. This ICode is publicly
available at \verb|https://github.com/wll745881210/MCIC|.
Please cite this paper if you use ICode.

\bibliographystyle{apj}
\bibliography{astro}

\begin{thebibliography}{}
\expandafter\ifx\csname natexlab\endcsname\relax\def\natexlab#1{#1}\fi

\bibitem[{{Blandford}(1999)}]{1999Blandford}
{Blandford}, R.~D. 1999, in Astronomical Society of the Pacific Conference
  Series, Vol. 182, Galaxy Dynamics - A Rutgers Symposium, ed. {D.~R.~Merritt,
  M.~Valluri, \& J.~A.~Sellwood}, 87--+

\bibitem[{{Blumenthal} \& {Gould}(1970)}]{1970blumenthal}
{Blumenthal}, G.~R., \& {Gould}, R.~J. 1970, Reviews of Modern Physics, 42, 237

\bibitem[{{Brocksopp} {et~al.}(2006){Brocksopp}, {Starling}, {Schady}, {Mason},
  {Romero-Colmenero}, \& {Puchnarewicz}}]{2006Brocksopp}
{Brocksopp}, C., {Starling}, R.~L.~C., {Schady}, P., {et~al.} 2006, \mnras,
  366, 953

\bibitem[{{Bryan} {et~al.}(2014){Bryan}, {Norman}, {O'Shea}, {Abel}, {Wise},
  {Turk}, {Reynolds}, {Collins}, {Wang}, {Skillman}, {Smith}, {Harkness},
  {Bordner}, {Kim}, {Kuhlen}, {Xu}, {Goldbaum}, {Hummels}, {Kritsuk}, {Tasker},
  {Skory}, {Simpson}, {Hahn}, {Oishi}, {So}, {Zhao}, {Cen}, {Li}, \& {The Enzo
  Collaboration}}]{2014Bryan}
{Bryan}, G.~L., {Norman}, M.~L., {O'Shea}, B.~W., {et~al.} 2014, \apjs, 211, 19

\bibitem[{{Burlon} {et~al.}(2011){Burlon}, {Ajello}, {Greiner}, {Comastri},
  {Merloni}, \& {Gehrels}}]{2011Burlon}
{Burlon}, D., {Ajello}, M., {Greiner}, J., {et~al.} 2011, \apj, 728, 58

\bibitem[{{Cen} {et~al.}(1995){Cen}, {Kang}, {Ostriker}, \& {Ryu}}]{1995Cen}
{Cen}, R., {Kang}, H., {Ostriker}, J.~P., \& {Ryu}, D. 1995, \apj, 451, 436

\bibitem[{{Di Matteo}(1998)}]{1998DiMatteo}
{Di Matteo}, T. 1998, \mnras, 299, L15

\bibitem[{{Draine}(2011)}]{2011Draine}
{Draine}, B.~T. 2011, {Physics of the Interstellar and Intergalactic Medium},
  ed. {Draine, B.~T.}

\bibitem[{{Elvis}(2000)}]{2000Elvis}
{Elvis}, M. 2000, \apj, 545, 63

\bibitem[{{Fabian} {et~al.}(1989){Fabian}, {Rees}, {Stella}, \&
  {White}}]{1989Fabian}
{Fabian}, A.~C., {Rees}, M.~J., {Stella}, L., \& {White}, N.~E. 1989, \mnras,
  238, 729

\bibitem[{{Falocco} {et~al.}(2012){Falocco}, {Carrera}, {Corral}, {Laird},
  {Nandra}, {Barcons}, {Page}, \& {Digby-North}}]{2012Falocco}
{Falocco}, S., {Carrera}, F.~J., {Corral}, A., {et~al.} 2012, \aap, 538, A83

\bibitem[{{Gallagher} {et~al.}(2002){Gallagher}, {Brandt}, {Chartas}, \&
  {Garmire}}]{2002Gallagher}
{Gallagher}, S.~C., {Brandt}, W.~N., {Chartas}, G., \& {Garmire}, G.~P. 2002,
  ApJ, 567, 37

\bibitem[{{Haardt}(1993)}]{1993haardt_b}
{Haardt}, F. 1993, \apj, 413, 680

\bibitem[{{Haardt} \& {Maraschi}(1991)}]{1991Haardt}
{Haardt}, F., \& {Maraschi}, L. 1991, \apjl, 380, L51

\bibitem[{{Haardt} \& {Maraschi}(1993)}]{1993Haardt}
---. 1993, \apj, 413, 507

\bibitem[{{Haardt} {et~al.}(1994){Haardt}, {Maraschi}, \&
  {Ghisellini}}]{1994Haardt}
{Haardt}, F., {Maraschi}, L., \& {Ghisellini}, G. 1994, \apjl, 432, L95

\bibitem[{{Haug}(1975)}]{1975Haug}
{Haug}, E. 1975, Zeitschrift Naturforschung Teil A, 30, 1546

\bibitem[{{Hopkins} \& {Quataert}(2010)}]{2010Hopkins}
{Hopkins}, P.~F., \& {Quataert}, E. 2010, \mnras, 407, 1529

\bibitem[{{Hopkins} \& {Quataert}(2011)}]{2011Hopkins}
---. 2011, \mnras, 415, 1027

\bibitem[{{Ichimaru}(1977)}]{1977Ichimaru}
{Ichimaru}, S. 1977, \apj, 214, 840

\bibitem[{{Igumenshchev} \& {Abramowicz}(2000)}]{2000Igumenshchev}
{Igumenshchev}, I.~V., \& {Abramowicz}, M.~A. 2000, \apjs, 130, 463

\bibitem[{{Kinzer} {et~al.}(1997){Kinzer}, {Jung}, {Gruber}, {Matteson},
  {Peterson}, \& {L.~E.}}]{1997Kinzer}
{Kinzer}, R.~L., {Jung}, G.~V., {Gruber}, D.~E., {et~al.} 1997, \apj, 475, 361

\bibitem[{{Liu} {et~al.}(2003){Liu}, {Mineshige}, \& {Ohsuga}}]{2003Liu}
{Liu}, B.~F., {Mineshige}, S., \& {Ohsuga}, K. 2003, \apj, 587, 571

\bibitem[{{Liu} {et~al.}(2002){Liu}, {Mineshige}, \& {Shibata}}]{2002Liu}
{Liu}, B.~F., {Mineshige}, S., \& {Shibata}, K. 2002, \apjl, 572, L173

\bibitem[{{Lusso} {et~al.}(2010){Lusso}, {Comastri}, {Vignali}, {Zamorani},
  {Brusa}, {Gilli}, {Iwasawa}, {Salvato}, {Civano}, {Elvis}, {Merloni},
  {Bongiorno}, {Trump}, {Koekemoer}, {Schinnerer}, {Le Floc'h}, {Cappelluti},
  {Jahnke}, {Sargent}, {Silverman}, {Mainieri}, {Fiore}, {Bolzonella}, {Le
  F{\`e}vre}, {Garilli}, {Iovino}, {Kneib}, {Lamareille}, {Lilly}, {Mignoli},
  {Scodeggio}, \& {Vergani}}]{2010Lusso}
{Lusso}, E., {Comastri}, A., {Vignali}, C., {et~al.} 2010, \aap, 512, A34

\bibitem[{{Mestel}(1963)}]{1963Mestel}
{Mestel}, L. 1963, \mnras, 126, 553

\bibitem[{{Miller} \& {Stone}(2000)}]{2000Miller}
{Miller}, K.~A., \& {Stone}, J.~M. 2000, \apj, 534, 398

\bibitem[{{Narayan} {et~al.}(1998){Narayan}, {Mahadevan}, \&
  {Quataert}}]{1998Narayan}
{Narayan}, R., {Mahadevan}, R., \& {Quataert}, E. 1998, in Theory of Black Hole
  Accretion Disks, ed. M.~A. {Abramowicz}, G.~{Bj{\"o}rnsson}, \& J.~E.
  {Pringle}, 148--182

\bibitem[{{Narayan} \& {Yi}(1994)}]{1994Narayan}
{Narayan}, R., \& {Yi}, I. 1994, \apjl, 428, L13

\bibitem[{{Ohsuga} \& {Mineshige}(2011)}]{2011Ohsuga}
{Ohsuga}, K., \& {Mineshige}, S. 2011, \apj, 736, 2

\bibitem[{{Pappa} {et~al.}(2001){Pappa}, {Stewart}, {Georgantopoulos},
  {Griffiths}, {Boyle}, \& {Shanks}}]{2001Pappa}
{Pappa}, A., {Stewart}, G.~C., {Georgantopoulos}, I., {et~al.} 2001, MNRAS,
  327, 499

\bibitem[{{Pietrini} \& {Krolik}(1995)}]{1995Pietrini}
{Pietrini}, P., \& {Krolik}, J.~H. 1995, \apj, 447, 526

\bibitem[{{Reynolds} \& {Nowak}(2003)}]{2003Reynolds}
{Reynolds}, C.~S., \& {Nowak}, M.~A. 2003, \physrep, 377, 389

\bibitem[{{Rybicki} \& {Lightman}(1979)}]{1979rybicki}
{Rybicki}, G.~B., \& {Lightman}, A.~P. 1979, {Radiative processes in
  astrophysics}

\bibitem[{{Sazonov} {et~al.}(2008){Sazonov}, {Krivonos}, {Revnivtsev},
  {Churazov}, \& {Sunyaev}}]{2008Sazonov}
{Sazonov}, S., {Krivonos}, R., {Revnivtsev}, M., {Churazov}, E., \& {Sunyaev},
  R. 2008, \aap, 482, 517

\bibitem[{{Shakura} \& {Sunyaev}(1973)}]{1973Shakura}
{Shakura}, N.~I., \& {Sunyaev}, R.~A. 1973, \aap, 24, 337

\bibitem[{{Spitzer}(1962)}]{1962Spitzer}
{Spitzer}, L. 1962, {Physics of Fully Ionized Gases}

\bibitem[{{Srednicki}(2007)}]{2007srednick}
{Srednicki}, M. 2007, {Quantum Field Theory}

\bibitem[{{Stone} \& {Norman}(1992)}]{1992StoneZeus}
{Stone}, J.~M., \& {Norman}, M.~L. 1992, \apjs, 80, 753

\bibitem[{{Sunyaev} \& {Titarchuk}(1980)}]{1980Sunyaev}
{Sunyaev}, R.~A., \& {Titarchuk}, L.~G. 1980, \aap, 86, 121

\bibitem[{{Tanaka} {et~al.}(1995){Tanaka}, {Nandra}, {Fabian}, {Inoue},
  {Otani}, {Dotani}, {Hayashida}, {Iwasawa}, {Kii}, {Kunieda}, {Makino}, \&
  {Matsuoka}}]{1995Tanaka}
{Tanaka}, Y., {Nandra}, K., {Fabian}, A.~C., {et~al.} 1995, \nat, 375, 659

\bibitem[{{Tomsick} {et~al.}(2009){Tomsick}, {Yamaoka}, {Corbel}, {Kaaret},
  {Kalemci}, \& {Migliari}}]{2009tomsick}
{Tomsick}, J.~A., {Yamaoka}, K., {Corbel}, S., {et~al.} 2009, \apjl, 707, L87

\bibitem[{{Vasudevan} \& {Fabian}(2009)}]{2009vasudevan}
{Vasudevan}, R.~V., \& {Fabian}, A.~C. 2009, \mnras, 392, 1124

\bibitem[{{Veledina} {et~al.}(2011){Veledina}, {Vurm}, \&
  {Poutanen}}]{2011Veledina}
{Veledina}, A., {Vurm}, I., \& {Poutanen}, J. 2011, \mnras, 414, 3330

\bibitem[{{Walter} \& {Courvoisier}(1992)}]{1992walter}
{Walter}, R., \& {Courvoisier}, T.~J.-L. 1992, \aap, 258, 255

\end{thebibliography}
\end{document}